\documentclass[11pt,a4paper]{article}

\usepackage{amsmath}
\usepackage{amssymb}
\usepackage{appendix}
\usepackage{multirow}
\usepackage{geometry}
\usepackage{float}
\usepackage{xcolor}
\usepackage{adjustbox}
\usepackage{tabularx}
\newcolumntype{L}{>{\raggedright\arraybackslash}X}
\newcolumntype{R}{>{\raggedleft\arraybackslash}X}
\newcolumntype{C}{>{\centering\arraybackslash}X}

\usepackage{cite}

\usepackage{graphicx}
\usepackage{float} 
\usepackage[caption=false]{subfig}
\usepackage[colorlinks=true]{hyperref}
\usepackage[capitalise]{cleveref}

\usepackage{cancel}

\newcommand{\eq}[1]{\begin{equation} #1 \end{equation}}

\newcommand{\abs}[1]{\left|#1\right|}
\newcommand{\av}[1]{\langle #1 \rangle}

\newcommand{\heff}{\mathcal{H}_{\rm eff}}

\newcommand{\op}{\mathcal{O}}

\newcommand{\Ceff}[1]{{\cal C}^{\rm eff}_{#1}}

\newcommand{\Cc}[1]{{\cal C}_{#1}}

\newcommand{\HVK}{{\bar h}_V}
\newcommand{\HAK}{{\bar h}_A}
\newcommand{\HSK}{{\bar h}_S}
\newcommand{\HPK}{{\bar h}_P}
\newcommand{\HTensK}{{\bar h}_T}
\newcommand{\HTenstK}{{\bar h}_{T_t}}

\newcommand{\HSKbar}{{\bar h}_S^*}
\newcommand{\HPKbar}{{\bar h}_P^*}

\newcommand{\HTenstKbar}{{\bar h}_{T_t}^*}

\newcommand{\KallenB}{\lambda_{B}}
\newcommand{\KallenBs}{\lambda_{B_s}}

\definecolor{nicered}{rgb}{0.7,0.1,0.1}
\definecolor{nicegreen}{rgb}{0.063,0.47,.58}

\begin{document}

\begin{flushright}

TUM-HEP-1275/20\\[0.2cm]
\today
\end{flushright}

\vspace{0.3cm}

\begin{center}
{\bf {\Large 
The time-dependent angular analysis of $B_d\to K_S\ell\ell$,\\
a new benchmark for new physics
}
}

\vspace{0.5cm}

S\'ebastien Descotes-Genon$^{a}$, Mart\'in Novoa-Brunet$^a$, K. Keri Vos$^b$

\vspace{0.2cm}

\emph{$^a$  Universit\'e Paris-Saclay, CNRS/IN2P3, IJCLab, 91405 Orsay, France}

\vspace{0.2cm}

\emph{$^b$ Physik Department, Technische Universit\"at M\"unchen, \\ James-Franck-Str.~1, D-85748 Garching, Germany}

\end{center}

\vspace{0.5cm}

\begin{abstract}
 We consider the time-dependent analysis of $B_d\to K_S\ell\ell$ taking into account the time-evolution of the $B_d$ meson and its mixing into $\bar{B}_d$. We discuss the angular conventions required to define the angular observables in a transparent way  with respect to CP conjugation. The inclusion of time evolution allows us to identify six new observables, out of which three could be accessed from a time-dependent tagged analysis. We also show that these observables could be obtained by time-integrated measurements in a hadronic environment if flavour tagging is available.
 We provide simple and precise predictions for these observables in the SM and in NP models with real contributions to SM and chirally flipped operators,
 which are independent of form factors and
 charm-loop contributions. As such, these observables
 provide robust and powerful cross-checks of the New Physics scenarios currently favoured by global fits to $b\to s\ell\ell$ data. In addition, we discuss the sensitivity of these observables with respect to NP scenarios involving scalar and tensor operators, or CP-violating phases. We illustrate how these new observables can provide a benchmark to discriminate among the various NP scenarios in $b\to s\mu\mu$. We discuss the extension of these results for $B_s$ decays into $f_0$, $\eta$ or $\eta'$.
\end{abstract}

\vspace{0.5cm}

\section{Introduction}

The $b\to s\ell^+\ell^-$ transitions have been the focus of an intense theoretical and experimental activity over the last few years. Indeed, this Flavour-Changing Neutral-Current transition is CKM and loop suppressed in the SM and therefore very sensitive to New Physics (NP). Processes involving $b\to s\mu^+\mu^-$ at the quark level have been measured by several experiments, showing a series of deviations from the SM in the branching ratios for $B\to K\mu^+\mu^-$~\cite{Aaij:2014pli}, $B\to K^*\mu^+\mu^-$~\cite{Aaij:2014pli,Aaij:2013iag,Aaij:2016flj}, $B_s\to \phi\mu^+\mu^-$~\cite{Aaij:2015esa} as well as for the optimised angular observables~\cite{Matias:2012xw,Descotes-Genon:2013vna} in $B\to K^*\mu^+\mu^-$~\cite{Aaij:2013aln,Aaij:2015oid,Abdesselam:2016llu,ATLAS:2017dlm,CMS:2017ivg,Aaij:2020nrf}.
The branching fraction of $B_s\to \mu\mu$ seems also below the SM expectations~\cite{CMS:2014xfa,Aaij:2017vad,Aaboud:2018mst}.
Moreover, the comparison of $b\to s\mu^+\mu^-$ and $b\to se^+e^-$ through the measurements of $R_K$~\cite{Aaij:2014ora,Abdesselam:2019lab}, $R_{K^*}$~\cite{Aaij:2017vbb} and $B\to K^*\ell^+\ell^-$ angular observables~\cite{Aaij:2015dea,Wehle:2016yoi} for several values of the dilepton invariant mass hint at a violation of lepton flavour universality (LFU).

These deviations from the SM can be explained in a consistent and economical way within a model-independent effective field theory approach.  They need only a few shifts in the Wilson coefficients describing short-distance physics, as could be expected from New Physics (NP) violating lepton flavour universality and coupling to muons but not (or marginally) to electrons~(see the updated results in Ref.~\cite{Alguero:2019ptt} and other works in Refs.~\cite{Descotes-Genon:2013wba,Descotes-Genon:2015uva,Capdevila:2017bsm,Aebischer:2019mlg,Ciuchini:2019usw,Alok:2019ufo,Arbey:2019duh,Kumar:2019qbv,Datta:2019zca,Bhattacharya:2019dot,Biswas:2020uaq}). The corresponding violation of LFU between muons and electrons is indeed significant, around $25\%$ of the SM value for the semileptonic operator ${\cal O}_{9\mu}$, with several scenarios showing an equivalent ability to explain the observed deviations~\cite{Bifani:2018zmi}.

It is thus of primary interest to confirm and constrain further the scenarios of New Physics in $b\to s\ell^+\ell^-$ transitions. On the theory side, an ongoing effort is carried out to sharpen the predictions on the 
hadronic contributions to these decays (form factors, charm-loop contributions).
On the experimental side, one can collect more data (as already done by LHCb, CMS and ATLAS), exploit different experimental environments (in particular Belle II) and add new observables (for instance LFU angular observables~\cite{Alguero:2019pjc}). An interesting example is provided by the recent consideration of $\Lambda_b$ decays as an additional probe of $b\to s\mu\mu$ transitions benefiting from different theoretical and experimental systematics,
such as $\Lambda_b\to\Lambda(\to p \pi)\mu\mu$\cite{Gutsche:2013pp,Boer:2014kda,Roy:2017dum, Das:2018iap,Das:2018sms,Blake:2017une,Bhattacharya:2019der} and $\Lambda_b\to \Lambda(1520)(\to pK)\mu\mu$~\cite{Descotes-Genon:2019dbw,Aaij:2019bzx,Das:2020cpv,Amhis:2020phx}.
If we remain in the domain of meson decays, it is possible to consider higher-mass resonances~\cite{Lu:2011jm,Gratrex:2015hna,Aaij:2016kqt,Dey:2016oun,Das:2018orb}, with the issue of determining the corresponding hadronic contributions appropriately~\cite{Descotes-Genon:2019bud}. 

Another way of building new observables has also been discussed in Ref.~\cite{Descotes-Genon:2015hea} by exploiting neutral $B$-meson mixing and considering time-dependent observables. The changes induced by mixing have been discussed in Refs.~\cite{Bobeth:2008ij,Descotes-Genon:2015hea} for
light vector resonances into CP eigenstates such as $B_d\to K^{*0}(\to K_S\pi^0)\mu\mu$ and $B_s\to \phi(\to KK)\mu\mu$. The general discussion of CP violation comparing time-integrated and time-dependent observables
sheds some light on the interest of the new observables
obtained in Ref.~\cite{Descotes-Genon:2015hea}:
they correspond to
CP violation in the interference between decay and mixing,
they contain additional information not present in time-integrated observables (in particular concerning CP-odd ``weak'' phases) and they are not sensitive to the same hadronic uncertainties. In the context of $B\to K^*\mu\mu$, they lift some of the degeneracies among (time-integrated) angular observables that prevent us from separating the contributions from various helicity amplitudes~\cite{Matias:2012xw}.

In this article, we are going to follow the steps of Ref.~\cite{Descotes-Genon:2015hea} to analyse the simpler case where a neutral meson decays into a (CP-eigenstate) (pseudo)scalar meson and a lepton pair. Although our formalism is general, we will consider mainly $B_d\to K_S\mu\mu$ for illustration, since the (time-integrated) angular analysis of this decay has already been performed by LHCb using 3 fb$^{-1}$ of integrated luminosity~\cite{Aaij:2014tfa}. We will see that a time-dependent analysis of this decay yields 6 new observables, out of which 3 are promising experimentally. The very simple structure of these observables will allow us to show that they are very well determined within the Standard Model and that deviations from SM expectations can be analysed to determine whether scalar and tensor contributions or NP phases are involved.

In Sec.~\ref{sec:amplanalysis}, we recall the angular analysis of $B\to K\ell\ell$ without mixing, i.e.~the charged case, highlighting the angular convention required to connect CP-conjugate modes and the status  of the hadronic inputs needed for the theoretical computation. In Sec.~\ref{sec:withmixing}, we extend the discussion to the neutral case with mixing, discussing the CP-parity of the final state and deriving the 6 new time-dependent observables that can be measured in principle. In Sec.~\ref{sec:observables}, we focus on three of these new observables which are very precisely determined in the SM and can be used to probe various NP hypotheses (scalar and tensor contributions, NP ``weak'' phases), before concluding in Sec.~\ref{sec:conclusions}. In a first appendix, we show that our conclusions are not affected by the choice of a model for charm-loop contributions. In a second appendix, we discuss the case of $B_s$ decays into $f_0,\eta$ or $\eta'$ mesons showing that similar observables can be defined and computed.

\section{Angular analysis of $B^\pm\to K^\pm\ell\ell$}\label{sec:amplanalysis}

\subsection{Amplitude analysis}\label{sec:amplcharged}

The $b\to s\ell\ell$ transitions are described by the usual weak effective theory (WET), with  SM operators plus (potentially) NP operators with a chirally-flipped, scalar or tensor structure~\cite{Altmannshofer:2008dz}:
\eq{
\heff = -\frac{4 G_F}{\sqrt{2}}
\bigg[
\lambda_u\,[\Cc1 (\op_1^c-\op_1^u) +
\Cc2 (\op_2^c-\op_2^u)]
+\lambda_t\sum_{i\in I} \Cc{i} \op_{i}
\bigg]\ ,
}
where $\lambda_q = V_{qb} V_{qs}^*$ and $I = \{1c,2c, 3,4,5,6, 8, 7, 7', 9\ell,9'\ell,10\ell,10'\ell,S\ell,S'\ell,P\ell,P'\ell,T\ell,T'\ell \}$.
In the following, we neglect doubly Cabibbo suppressed contributions, with relative size of $O(\lambda^2)\simeq 4\%$ with $\lambda$ the usual parameter of the Wolfenstein parametrisation of the Cabibbo-Kobayashi-Maskawa (CKM) matrix, which leads us to neglect the contributions proportional to $\lambda_u$  in $\heff$.
The operators $\op_{1,..,6}$ and $\op_8$ are hadronic operators of the type
$(\bar s \Gamma b)(\bar q \Gamma' q)$ and
$(\bar s \gamma^{\mu\nu} T_a P_R b) G_{\mu\nu}^a$ respectively. These operators are not likely to receive very large contributions
from NP, as these would show up in non-leptonic $B$ decay amplitudes (see Refs.~\cite{Brod:2014bfa,Lenz:2019lvd,Jager:2019bgk} for a discussion of the room left for NP in these operators).
The main operators of interest
$\op_{7^{(\prime)},9^{(\prime)},10^{(\prime)},S^{(\prime)},P^{(\prime)},T^{(\prime)}}$ are given by:
\begin{align}
{\cal O}_{7^{(\prime)}} &= \frac{e}{(4\pi)^2} m_b [\bar{s} \sigma^{\mu\nu} P_{R(L)} b] F_{\mu\nu} \;, &
{\op}_{S^{(\prime)}\ell}  &= \frac{e^2}{(4\pi)^2} [\bar{s} P_{R(L)} b][\bar{\ell}\,\ell]\;, \nonumber
\\
\label{effops}
{\cal O}_{9^{(\prime)}\ell}  &=  \frac{e^2}{(4\pi)^2} [\bar{s} \gamma^\mu P_{L(R)} b] [\bar{\ell} \gamma_\mu \ell] \;, &
{\op}_{P^{(\prime)}\ell}  &= \frac{e^2}{(4\pi)^2} [\bar{s} P_{R(L)} b][\bar{\ell}\gamma_5\ell]\;, \\ \nonumber
{\cal O}_{10^{(\prime)}\ell}  &=  \frac{e^2}{(4\pi)^2} [\bar{s} \gamma^\mu P_{L(R)} b] [\bar{\ell} \gamma_\mu \gamma_5 \ell]\;, &
{\op}_{T^{(\prime)}\ell}  &=  \frac{e^2}{(4\pi)^2}
[\bar{s} \sigma_{\mu\nu}P_{R(L)} b][\bar{\ell}\sigma^{\mu\nu}\ell]\;.
\end{align}
In the SM, and at a scale $\mu_b=\op(m_b)$, the only non-negligible Wilson coefficients concerning the operators of
Eq.~(\ref{effops}) are $\Cc7^\text{SM}(\mu_b)\simeq -0.3$,
$\Cc9^\text{SM}(\mu_b)\simeq 4.1$ and $\Cc{10}^\text{SM}(\mu_b)\simeq -4.3$ -- the precise values are given in Table~\ref{tab:inputs}
and are identical for $\ell=e$ and $\ell=\mu$ due to the universality of lepton couplings in the SM.
All the Wilson coefficients might be affected by NP contributions, which can also violate Lepton Flavour Universality and be different for $\ell=e$ and $\ell=\mu$. For simplicity, in the following, we will omit the index $\ell$ if the context is clear enough to determine whether we consider a generic lepton or the specific case $\ell=\mu$.

Contributions from the semileptonic operators are factorizable and their matrix elements
can be written as
\eq{
\av{K\ell\ell| \op_\text{sl} |B} = \av{K|\Gamma^A |B} \av{\ell\ell|\Gamma'_A|0}\ ,
}
where $A$ denotes a collection of Lorentz indices and $\Gamma,\Gamma'$ are Dirac matrices. It is clear that all hadronic, dipole, and
semileptonic contributions can be recast as decays of the form
\eq{
B\to K N(\to \ell^+\ell^-)\ ,
\label{eq:BPN}
}
where $N$ has the quantum numbers of a boson, whose coupling pattern is determined by the
operators arising in the effective Hamiltonian.
In the SM, the structure of $\op_7,\op_9,\op_{10}$ shows that $N$ are spin-1 particles, coupling to both left- and right-handed fermions. This is in agreement with the presence of $\gamma^*$ and $Z$ penguin contributions, but it is also able to reproduce the contribution from box diagrams involving two $W$ bosons and a neutrino ($(V-A)(V-A)$ structure in the SM). In an extension of the SM yielding scalar (tensor) operators, one should add $N$ bosons with spin 0 (spin 2 respectively)~\cite{Gratrex:2015hna}. 

We can exploit Ref.~\cite{Gratrex:2015hna} in order to extract information starting with the charged decay. The angular distribution for $B^-\to K^-\ell\ell$ is
\begin{equation}
\frac{d^2 \Gamma(B^-\to K^-\ell\ell)}{dq^2 \; d\cos\theta_\ell} =   \bar{G}_{0}(q^2) + \bar{G}_{1}(q^2) \cos\theta_\ell+ \bar{G}_{2}(q^2)  \frac{1}{2}(3\cos^2\theta_\ell-1)= \sum_{i=0,1,2} \bar{G}_i(q^2) P_i(\cos\theta_\ell)\end{equation}
where $P_i$ denotes the $i$-th Legendre polynomial
in terms of the angle $\theta_\ell$ describing the emission of one of the charged leptons (its precise definition will be discussed in the following)
and $q=p_B-p_K$ is the momentum transfer. We have
\begin{eqnarray}
\label{eq:gi}
\bar{G}_{0} &=& \frac{4}{3} \left(1 + 2\hat{m}_\ell^2 \right) \left| \HVK \right|^2 +  \frac{4}{3}  \beta_\ell^2 \left| \HAK \right|^2  \nonumber + 2  \beta_\ell^2 \left| \HSK \right|^2   +2   \left| \HPK \right|^2  \nonumber \\
&& + \frac{8}{3} \left( 1 + 8 \hat{m}_\ell^2  \right) \left| \HTenstK \right|^2 + \frac{4}{3} \beta_\ell^2\left| \HTensK \right|^2   + 16 \hat{m}_\ell \,{\rm Im} \left[ \HVK \HTenstKbar \right]   \; , \nonumber \\
\bar{G}_{1} &=&  - 4  \beta_\ell \left(2  \hat{m}_\ell   \, {\rm Re} \left[ \HVK \HSKbar  \right] - {\rm Im} \left[2  \HTenstK \HSKbar  + \sqrt{2} \HTensK \HPKbar\right]  \right) \; , \nonumber \\
\bar{G}_{2} &=& - \frac{4  \beta_\ell^2 }{3} \left( \left| \HVK \right|^2 + \left| \HAK \right|^2 - 2 \left| \HTensK \right|^2 - 4 \left|\HTenstK \right|^2\right)\;, 
\end{eqnarray}
where we have used the notation:
\begin{equation}
    \hat{m}_\ell = \frac{ m_\ell}{\sqrt{q^2}}\,,
    \qquad
    \beta_\ell=\sqrt{1-4\hat{m}_\ell^2}\,.
\end{equation}

The matrix elements relevant to $\bar B \to \bar K$ transition yield the following form factors in the standard parametrisation:
\begin{eqnarray}
\langle K^-(p) |\bar{s} \gamma_\mu  b|  B^-(p_B)\rangle & =& \left ( p_B + p \right )_\mu f_{+} (q^2) + \frac{m_B^2 - m_K^2}{q^2}q_\mu \left( f_0 (q^2) - f_{+} ( q^2 ) \right) \; , \nonumber \\
\langle K^-(p) |\bar{s}  \sigma_{\mu \nu}  b| B^-(p_B)\rangle & =& i \left[ \left ( p_B + p \right )_\mu q_\nu -  \left ( p_B + p \right )_\nu q_\mu \right] \frac{f_T (q^2)}{m_B + m_K}  \; , \nonumber \\
\langle K^-(p) |\bar{s}    b| B^-(p_B)\rangle & =& \frac{m_B^2 - m_K^2 }{m_b - m_s} f_0 (q^2) \; ,
\end{eqnarray}
We find
\begin{eqnarray}
\HVK &=& {\cal N}\frac{\sqrt{\KallenB} }{2 \sqrt{q^2}} \left(\frac{2 m_b}{ m_B + m_K }  ( {\cal C}_{7} + {\cal C}_{7'}) f_T + ( {\cal C}_{9} + {\cal C}_{9'})   f_+  \right) \; ,  \nonumber \\
\HAK &=& {\cal N}\frac{\sqrt{\KallenB}  }{2 \sqrt{q^2}} ( {\cal C}_{10} + {\cal C}_{10'})  f_+ \; , \nonumber \\
\HSK &=& {\cal N} \frac{m_B^2-m_K^2}{2}\left(\frac{ ( {\cal C}_{S} + {\cal C}_{S'})}{m_b - m_s} \right)f_0 \; ,  \nonumber \\
\HPK &=&  {\cal N}\frac{m_B^2-m_K^2}{2}\left(\frac{ ( {\cal C}_{P} + {\cal C}_{P'})}{m_b - m_s} + \frac{ 2m_\ell}{ q^2  }   ( {\cal C}_{10} + {\cal C}_{10'}) \right)f_0 \; ,  \nonumber \\
\HTensK &=&  - i {\cal N}\frac{\sqrt{\KallenB}  }{\sqrt{2} \left( m_B + m_K \right)} \left({\cal C}_{T} - {\cal C}_{T'} \right) f_T \; ,  \nonumber \\
\HTenstK &=& - i {\cal N}\frac{\sqrt{\KallenB}  }{2 \left( m_B + m_K \right)} \left({\cal C}_{T} + {\cal C}_{T'} \right) f_T \; , 
\end{eqnarray}
with the normalisation factor ${\cal N}$,
\begin{equation}
\label{eq:NN}
{\cal N}=  -\frac{\alpha G_F}{\pi} V_{ts}^*V_{tb}\sqrt{\frac{q^2\beta_\ell\sqrt{\KallenB}}{2^{10} \pi^3 m_B^3}}  \;,
\end{equation}
where $\KallenB \equiv \lambda(m_B^2, m_K^2, q^2) $ (with $\lambda(a,b,c)$ is the K\"all\'en-function) is related to 
the absolute value of the three-momentum of the $K^*$. Note that the normalisation factor ${\cal N}$ disagrees with (the square root of) the normalisation factor of Ref.~\cite{Gratrex:2015hna} by a factor $2\sqrt{2}$, but it is in agreement with Refs.~\cite{Bobeth:2007dw,Becirevic:2012fy}.

Following Ref.~\cite{Gratrex:2015hna},
we use the LHCb conventions for the charged case,  so that $\theta_\ell$ is defined as the angle between the
$\ell^-$ three-momentum and the opposite of the $B^-$ three-momentum in the dilepton rest frame in the case of $B^-\to K^-\ell\ell$, but $\ell^+$ and $B^+$ in the case of $B^+\to K^+\ell\ell$. With this convention,
$d^2 \Gamma(B^+\to K^+\ell\ell)/(dq^2 \; d\cos\theta_\ell)$
has the same expression as $d^2\Gamma(B^-\to K^-\ell\ell)/(dq^2 \; d\cos\theta_\ell)$ above,
up to the replacement of the angular coefficients $\bar{G}$ depending on $\bar{h}$ by $G$ depending on $h$. The $h$ amplitudes are obtained from the $\bar{h}$ amplitudes by performing a complex conjugation of all the weak phases (this applies to ${\cal N}$ but also to the weak phases in the Wilson coefficients in the case of CP-violating New Physics). On the other hand, strong phases, in particular those stemming from charm loops generated by the four-quark operators and combining with ${\cal C}_9$ in the expressions of the angular observables, are the same in $h$ and $\bar{h}$. If all CP violating effects are neglected, one gets $G_i = \bar G_i$. 
 
\subsection{Hadronic inputs} 
 \label{sec:hadronicinputs}

In order to compute the amplitudes and the angular observables defined above, we need hadronic inputs for $f_{+,0,T}$. We may use the form factors obtained in Ref.~\cite{Khodjamirian:2017fxg} (for $f_+$ and $f_T$) and Ref.~\cite{Gubernari:2018wyi} (for $f_+$, $f_0$ and $f_T$). Both perform 
light-cone sum rules determinations at low $q^2$, using sum rules based on light-meson and $B$-meson distribution amplitudes, respectively. The authors of
Ref.~\cite{Gubernari:2018wyi} combine their results with
lattice QCD determination at high $q^2$ (coming from Ref.~\cite{Lattice:2015tia}). The observables built as ratios of angular coefficients $G_i$ depend actually on the ratios of form factors $f_0/f_+$ and $f_T/f_+$. It turns out that $f_T/f_+$ has little $q^2$-dependence and is very close to 1 within the uncertainties quoted, in agreement with 
the earlier discussion in Ref.~\cite{Becirevic:2012fy}
and with the expectations at low $q^2$ (large $K$ recoil) from Soft-Collinear Effective Theory~\cite{Charles:1998dr,Beneke:2000wa} and at high $q^2$ (low $K$ recoil) from Heavy Quark Effective Theory~\cite{Bobeth:2011nj}.
On the other hand $f_0/f_+$ has a linear dependence on $q^2$, so that a noticeable $q^2$ dependence of ratios of angular observables $G_i$ could be the sign of significant scalar/pseudoscalar contributions. 

\begin{figure}[t]
    \centering
    \includegraphics{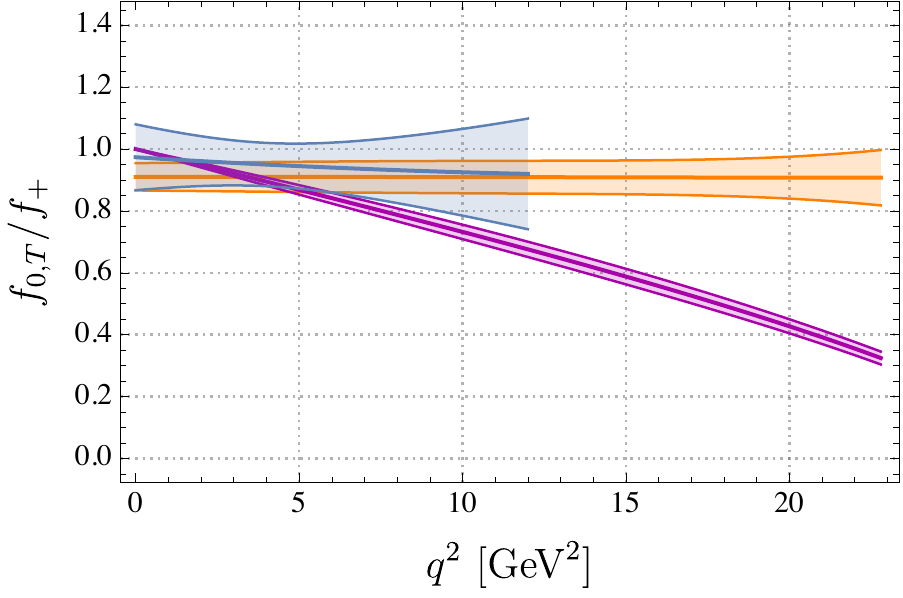}
    \caption{$f_0/f_+$ (purple) and $f_T/f_+$ (orange) form factors ratios for $B\to K$ with a lattice QCD and LCSR combination from Ref.~\cite{Gubernari:2018wyi} and $f_T/f_+$ (blue) coming from Ref.~\cite{Khodjamirian:2017fxg}.} 
    \label{fig:FF_Ratios}
\end{figure}

We have not explicitly indicated the contribution from $c\bar{c}$ loops which adds a $q^2$-dependent contribution to the coefficient ${\cal C}_9$, which features both a real and an imaginary part coming from strong phases. The size of this contribution has been under discussion for $B\to K^*\ell\ell$ (see Ref.~\cite{Capdevila:2017ert} and references therein for a recent discussion). For $B\to K\ell\ell$ at large recoil, the current estimates from light-cone sum rules~\cite{Khodjamirian:2010vf,Khodjamirian:2012rm} suggest a contribution of order 10\% percent of ${\cal C}_9$, with a moderate dependence on the dilepton invariant mass $q^2$. For instance, within $[1,6]\ {\rm GeV}^2$, the range of variation remains within:
\begin{equation}\label{eq:C9ccLCSR}
    {\rm LCSR\ contrib.\ for}\ q^2\in [1,6]\ {\rm GeV}^2: {\rm Re}\ {\cal C}_9^{c\bar{c}}=-0.26\pm 0.10\,, \qquad {\rm Im}\ {\cal C}_9^{c\bar{c}}=-0.49\pm0.27\,.
    \end{equation}
where we set the central value and range to cover the values from Ref.~\cite{Khodjamirian:2012rm}. These results can be compared with the results obtained by considering only the perturbative part of te $c\bar{c}$ contribution, for instance
\begin{eqnarray}\label{eq:C9ccpert}
    {\rm Perturbative\ contrib.\ at}\ q^2=1\ {\rm GeV}^2&:& {\rm Re}\ {\cal C}_9^{c\bar{c}}=0.16, \qquad {\rm Im}\ {\cal C}_9^{c\bar{c}}=0.17\,.\\
    q^2=6\ {\rm GeV}^2&:& {\rm Re}\ {\cal C}_9^{c\bar{c}}=0.11, \qquad {\rm Im}\ {\cal C}_9^{c\bar{c}}=0.17\,.
\end{eqnarray}
to which we do not attempt to assign a meaningful theoretical uncertainty.

These results are similar in size ($\simeq 10\%$) to the expected impact of charmonium resonances at low $K$-recoil of a few percent based on quark-hadron duality~\cite{Beylich:2011aq}. They are also in line with the dimensional estimates based on the $1/m_b$ suppression of these contributions. In the following, 
we will take the following estimate for the SM value of ${\cal C}_9$ including the effect of $c\bar{c}$ resonances both at low and large $K$-recoil:
\begin{equation}\label{eq:C9ccmodel}
    {\rm Our\ estimate}:\ {\cal C}_9^{\rm SM}={\cal C}_9^{\rm WET} (1+\rho e^{i\phi})\,,\qquad \rho\in[0,0.1]\,,\qquad \phi \in [0,2\pi]\,,
\end{equation}
where ${\cal C}_9^{\rm WET}$ corresponds to the outcome of the WET computation given in Table~\ref{tab:inputs}.
This simple order-of-magnitude estimate does not include any $q^2$-dependence, as would be expected for a proper description of the $c\bar{c}$-loop contributions~\cite{Capdevila:2017ert}. The alternative estimates Eqs.~(\ref{eq:C9ccLCSR}) and  (\ref{eq:C9ccpert}) will be used only to check that our results depend only very mildly on the model used for charm-loop contribution. Obviously, Eq.~(\ref{eq:C9ccmodel}) does not hold in the charmonium resonance region, where the $c\bar{c}$ pair becomes resonant and yields contributions that are much larger~\cite{Lyon:2014hpa}.

\subsection{Observables}\label{sec:angobstimeindep}

The angular observables $G_i$ can be recast into more traditional forms.
In addition to the decay rate and the forward-backward asymmetry, a third observable can be built from the $B^\pm\to K^\pm\ell\ell$ angular analysis~\cite{Bobeth:2007dw}. The corresponding CP-averaged observables  have the following expressions in terms of the angular coefficients:
\begin{equation}
    \Gamma_\ell=G_0+\bar{G}_0\,, \qquad
    A^\ell_{FB}=\frac{G_1+\bar{G}_1}{2(G_0+\bar{G}_0)}\,,\qquad
    F_H^\ell=1+\frac{G_2+\bar{G}_2}{G_0+\bar{G}_0}\,,
\end{equation}
leading to 
\begin{equation}
\frac{d^2 \Gamma(B^-\to K^-\ell\ell)}{dq^2 \; d\cos\theta_\ell}+
\frac{d^2 \Gamma(B^+\to K^+\ell\ell)}{dq^2 \; d\cos\theta_\ell}
=   2\Gamma_\ell\left[\frac{1}{2}F_H^\ell +A_{FB}^\ell\cos\theta_\ell+\frac{3}{4}(1-F_H^\ell)(1-\cos^2\theta_\ell)\right]\,.
\end{equation}

As can be seen from the above equations, in the absence of tensor and scalar contributions and neglecting $\hat{m}_\ell$ corrections which are relevant only at very low $q^2$, one has the simple relations
\begin{equation} \label{eq:approxSM}
    \bar{G}_0=-\bar{G}_2+2|\bar{h}_P|^2\simeq -\bar{G}_2\,, \qquad \bar{G}_1=0
\end{equation}
(and the same for $G_i$). The observable $F_H^\ell$ is proportional to $(G_0+\bar{G}_0)+(G_2+\bar{G}_2)$, and thus probes the first relation in Eq.~(\ref{eq:approxSM}).
A non-vanishing value of $F_H^\ell$ can be attributed to NP in tensor and/or scalar contributions. On the other hand,
a non-vanishing $A^\ell_{FB}$, related to $G_1+\bar{G}_1$, 
probes the second relation in Eq.~(\ref{eq:approxSM}) and
would be a clear signal of New Physics from scalar or tensor contributions, but we can see from Eq.~(\ref{eq:gi}) that they should correspond to very large scalar contributions (to beat the $m_\ell$-suppressing factors) and/or to (pseudo)tensor and (pseudo)scalar contributions.

One can also think of building CP-violating observables of the form
\begin{equation}\label{eq:AiCPviolating}
    A_i=\frac{G_i-\bar{G}_i}{G_i+\bar{G}_i}\,.
\end{equation}
Neglecting again $m_\ell$-suppressed contributions, we see that these observables probe differences of the form $|h_X|^2-|\bar{h}_X|^2$, which vanish unless both strong and weak phases are present. 
Assuming that NP contributions do not yield any significant strong phases in the short-distance Wilson coefficients,
it can be easily seen that only $h_V$ involves strong phases (due to the $c\bar{c}$-loops) and thus only the presence of CP-violating NP phases in ${\cal C}_{7,7'}$ and ${\cal C}_{9,9'}$ can be probed by these observables. 

In Appendix~\ref{app:predictions} we provide predictions for these observables in the SM and in a few NP scenarios. It is quite clear that they yield rather similar central values in all scenarios, with hadronic uncertainties that are rather large compared to the sensitivity to NP contributions. This makes the NP interpretation of deviations in the measurement of these observables rather challenging.

As a conclusion, the CP-averaged observables built from the angular analysis of $B^+\to K^+\ell\ell$ have interesting abilities to probe scalar and tensor NP contributions, but if deviations from the SM are observed, these observables are not sufficient to pin down the actual source of the contributions. The CP asymmetries associated with the same observables probe only the presence of NP phases in a limited subset of Wilson coefficients.

\begin{table}[t]
	\begin{center}
\begin{tabularx}{0.98\textwidth}{|c|c|c|c|C|C|C|}
\hline\multicolumn{7}{|c|}{$B_d\to K_{S(L)} \ell^+\ell^-$ parameters}  \\ \hline
$\eta(K_S)$ &$\eta(K_L)$ & $\phi$  & $\Delta \Gamma$ & $x=\Delta m/\Gamma$ & $y=\Delta\Gamma/(2\Gamma)$&$\tau_{B_d} [ps]$\\
\hline
 1& -1 & $-2\beta$ & $\simeq 0$ & $0.769\pm0.004$ & $0.0005\pm0.005$&$1.519\pm0.004$\\ \hline
\end{tabularx}
\vskip 1pt
\begin{tabularx}{0.98\textwidth}{|c|c|c|c|C|C|C|}
\hline
\multicolumn{7}{|c|}{$B_s\to f_0(\eta,\eta') \ell^+\ell^-$ parameters}  \\ \hline
$\eta(f_0)$ & $\eta(\eta,\eta')$ & $\phi$ &  $\Delta \Gamma$ & $x=\Delta m/\Gamma$ & $y=\Delta \Gamma/(2\Gamma)$&$\tau_{B_s} [ps]$\\
\hline 1 & -1 & $2\beta_s$ &  $\neq 0$ & $26.81\pm 0.08$ & $0.0675\pm 0.004$&$1.515\pm0.004$\\
\hline
\end{tabularx}
\vskip 1pt

\begin{tabularx}{0.98\textwidth}{|C|c|c|c|c|}
\hline
\multicolumn{5}{|c|}{CKM parameters}  \\ \hline
$\sin(-2\beta)$ & $\sin(2\beta_s)$ &
${\rm Re}[V_{ts}] $ &
  ${\rm Im}[V_{ts}] \cdot 10^{3}$& $V_{tb}$\\ \hline
  $-0.71\pm0.01$ & $0.0371\pm0.0007$ &
$ -0.0407\pm 0.0004$ &
  $-0.75 \pm 0.02$  & $0.99913\pm 0.00002$\\
\hline
\end{tabularx}

\vskip -1pt
\begin{tabularx}{0.98\textwidth}{|C|C|C|C|}
$\lambda$  & $A$ & $\bar\rho$ & $\bar{\eta}$ \\
\hline
 $0.22493\pm 0.00016$&  $0.819\pm 0.010$ & $0.159\pm 0.008$ & $0.351\pm
   0.007$ \\
\hline
\end{tabularx}
\vskip 1pt

	\begin{tabularx}{0.98\textwidth}{|C|C|C|}
	\hline
	\multicolumn{3}{|c|}{Masses [GeV]}\\
		\hline 
$\bar{m_b}(m_b)$  &  $\bar{m}_s(m_b)$ & $m_\mu $ 
      \\ \hline
      $4.18\pm0.03$  & $ 0.078 \pm 0.007$  & $ 0.106$
      \\ \hline\hline
	\multicolumn{3}{|c|}{Wilson Coefficients at $\mu=m_b$}\\
		\hline 
  $\Ceff7$ & $\Cc9$    &    $\Cc{10} $ 
      \\ \hline
  $-0.292$ & $4.07$    &    $ -4.31$ 
      \\ \hline
		\end{tabularx}
		\caption{Input parameters used to determine the SM predictions. Decay parameters are from Ref.~\cite{PDG2019}. The CKM values are obtained from the symmetrised confidence intervals for the Wolfenstein parameters $\lambda,A,\bar\rho,\bar\eta$ given in Ref.~\cite{CKMfitter}, while for $m_b$ and $m_s$ we use the $\overline{\rm{MS}}$ masses at $m_b$~\cite{PDG2019}. The SM Wilson coefficients are from Ref.~\cite{Descotes-Genon:2013vna}. The form factors (not recalled here) are taken from Ref.~\cite{Gubernari:2018wyi}.}
		\label{tab:inputs}
	\end{center}
\end{table}

\section{Angular analysis of $B_d\to K_S\ell\ell$} \label{sec:withmixing}

\subsection{From the charged case to the neutral one}
\label{sec:neutralcase}

Before analysing the impact of time evolution and mixing, we must first determine how the above formulae must be adapted to the neutral case if mixing were neglected.  We must perform the changes in the equations of Sec.~\ref{sec:amplcharged}:
\begin{equation}
   B^-\to \bar{B}_d\,,
   \qquad B^+ \to B_d\,,
   \qquad K^-\to \bar{K}^0\,,
   \qquad K^+\to K^0\,.
\end{equation}
We have then 
to consider the CP-states rather than the flavour states for the kaon with the following phase convention:
\begin{equation}
|K^0\rangle \sim d\bar{s}\,, \qquad |\bar{K}^0\rangle \sim s\bar{d}\,, \qquad
|K^+\rangle \sim u\bar{s}\,, \qquad |K^-\rangle \sim s\bar{u}\,,
\end{equation}
so that 
\begin{equation}
    |K_S\rangle \sim \frac{|K^0\rangle+|\bar{K}^0\rangle}{\sqrt{2}}\,,\qquad
    |K_L\rangle \sim \frac{|K^0\rangle-|\bar{K}^0\rangle}{\sqrt{2}}\,,
\end{equation}
where we have neglected the small amount of 
%\mn{indirect and} direct 
CP violation in the kaon system %decays 
and nition $CP|K^0\rangle=|\bar{K}^0\rangle$
(and similarly $CP|B_d\rangle=|\bar{B}_d\rangle$
The expressions for $\bar{h}_X(\bar{B}_d\to K_S\ell\ell)$  and
$h_X(B_d\to K_S\ell\ell)$
are obtained
from $\bar{h}_X(\bar{B}_d\to \bar{K}^0\ell\ell)$ and
$h_X(B_d\to K^0\ell\ell)$
by dividing the normalisation 
${\cal N}$ by $\sqrt{2}$ in both cases~\footnote{
For $K_L$, we would obtain the amplitudes by dividing the normalisation ${\cal N}$ by $-\sqrt{2}$ and by
$\sqrt{2}$, respectively.}. The latter are equal to the charged amplitudes described in the previous section in the isospin limit, so that we have
\begin{equation}
\bar{h}_X(\bar{B}_d\to K_S\ell\ell)=\frac{1}{\sqrt{2}}\bar{h}_X(B^-\to K^-\ell\ell)
\qquad 
h_X(B_d\to K_S\ell\ell)=\frac{1}{\sqrt{2}}h_X(B^+\to K^+\ell\ell)
\end{equation}

Following Ref.~\cite{Descotes-Genon:2015hea},
the discussion of $B_d\to K_S\ell\ell$ requires the same convention for both $B_d$ and $\bar{B}_d$, since the decay is not flavour specific. Before taking into account mixing, and following the arguments of Ref.~\cite{Descotes-Genon:2015hea} that we will discuss extensively below, we define $\theta_\ell$ as the angle between $\ell^-$ and $K_S$ (similarly to the case of $B^+\to K^+\ell\ell$) for both $B_d$ and $\bar{B}_d$ decays. This yield.
\begin{eqnarray}
\frac{d^2\Gamma[B_d\to K_S\ell^+\ell^-]}
{ds\  d\!\cos\theta_\ell\ }&=&
\sum_i G_i(s) P_i(\cos\theta_\ell)\,,
\label{Gamma}\\
\frac{d^2\Gamma[\bar{B}_d\to K_S\ell^+\ell^-]}
{ds\  d\!\cos\theta_\ell\ }&=&
\sum_i\zeta_i\bar{G}_i(s) P_i(\cos\theta_\ell)\,,
\label{Gammabar}
\end{eqnarray}
where $\zeta_{0,2}=1$ and $\zeta_1=-1$ and the $G_i$ ($\bar{G}_i$) are defined in terms of $h_X(B_d\to K_S)$ ($\bar{h}_X(\bar{B}_d\to K_S)$). In the absence of CP violation, we would have $G_i=\bar{G}_i$.

We stress that Eqs.~(\ref{Gamma})-(\ref{Gammabar}) arise just from the identification of kinematics of CP-conjugate decays, and
do not rely on any intrinsic CP-parity of the initial or final states involved. We will see now that this choice of conventions is justified by the analysis of the properties of the amplitudes under CP conjugation.

\subsection{CP-parity of the final state}

We now turn to the case of decays into CP eigenstates: $B\to f_{CP}$. In this context, it is useful to define two
different angular coefficients $\widetilde G_i$, $\bar G_i$ which are CP conjugates of $G_i$:
\begin{itemize}
\item the angular coefficients $\widetilde{G}_i$ formed by replacing $A_X$ by $\widetilde A_X\equiv A_X(\bar{B}_d\to f_{CP})$
(without CP-conjugation applied on $f_{CP}$), which appear naturally in the study of time evolution due to mixing,
where both $B$ and $\bar B$ decay into the same final state $f_{CP}$.
\item the angular coefficients $\bar{G}_i$, obtained by considering $\bar A_X\equiv A_X(\bar B_d \to \overline f_{CP})$ (with CP-conjugation applied to $f_{CP}$), which can be obtained from $A_X$ by changing the sign of all weak phases, and arise naturally when discussing CP violation from the theoretical point of view.
\end{itemize}
In the case of interest, we have to consider 
the transversity amplitudes:
\eq{
\bar A_X \equiv A_X(\bar B \to \bar M \ell^+\ell^-)\ ,\quad
\widetilde A_X \equiv A_X(\bar B \to M\ell^+\ell^-)\ ,
}
where $X=V,A,S,P,T,T_t$, and we have $\bar{A}_X=\bar{h}_X$.
These two sets of amplitudes are related by
\eq{
\widetilde A_X = \eta_X \bar A_X\,,
}
where $\eta_X$ are the CP-parities associated to the different transversity amplitudes. 
We follow the arguments of Refs.~\cite{Dunietz:1990cj,Descotes-Genon:2015hea} in order to determine the value of $\eta_X$. Adapting the arguments of Ref.~\cite{Dunietz:1990cj} to the decay $B\to MN$, where $M$ is stable (under the strong interaction) and $N$ decays into the dilepton pair, leads to
\begin{equation}
    \eta_X=\eta(M)\eta(N)(-1)^{\tau(N)}\,,
\end{equation}
where $M=K_S$ here. The assignment of the CP-parity $\eta(N)$ and the transversity $\tau(N)$ requires some discussion. 

Concerning the CP-parity $\eta(N)$, we can start from the helicity amplitude analysis performed in Ref.~\cite{Gratrex:2015hna}, associating the lepton matrix  elements 
$\langle \ell^-(\lambda_1)\ell^+(\lambda_2)|\bar\ell \Gamma^X \ell|0\rangle$
to the amplitudes $h_X$:
\begin{equation}
\Gamma^S=1,\qquad \Gamma^P=\gamma^5\,, \quad
\Gamma^V=\gamma^\mu\omega_\mu(\lambda)\,, \quad
\Gamma^A=\gamma^\mu\gamma^5\omega_\mu(\lambda)\,, \quad
\Gamma^T=\sigma^{\mu\nu}\omega^{1\lambda}_{\mu\nu}\,, \quad
\Gamma^{T_t}=\sigma^{\mu\nu}\omega^{t\lambda}_{\mu\nu}\,,
\end{equation}
where $\lambda=\lambda_1-\lambda_2$ (equal to $-1,0$ or 1). The polarisation vectors $\omega_\mu(\lambda)$ form the usual basis for $\lambda=t,0,+1,-1$, with $\omega_\mu(t)=q_\mu/\sqrt{q^2}$.
The rank-2 polarisation tensors $\omega^{J\lambda}_{\mu\nu}$
are less familiar objects, but they correspond to products of  polarisation vectors. On one hand, we have  $\omega^{t\lambda}_{\mu\nu}=\omega_\mu(t)\omega_\nu(\lambda)$
and on the other hand $\omega^{1\lambda}_{\mu\nu}$ is a linear combination of products of polarisation vectors $\omega_\mu(\lambda_1)\omega_\nu(\lambda_2)$ with $\lambda_1$ and $\lambda_2$ being either $0,-1$ or $1$, but not timelike. This formulation allows us to determine the parity $\eta(N)$: since we
assume that CP-parity is conserved
through the decay, we can determine the CP-parity of $N$ through that of the lepton matrix element it couples to, taking into account the sign difference between the time-like polarisation and the space-like polarisations. The corresponding parities of the fermion bilinears with different Dirac matrices can easily be found in the discussion of the Dirac algebra in textbooks on quantum field theory, for instance Ref.~\cite{Peskin:1995ev}.

Concerning the transversity $\tau(N)$, we can then use the following two  statements: first, the helicity of $N$ is $\lambda(N)=0$ since both $B$ and $M$ are spin-0 mesons, and second, the antisymmetric structure of the tensor operators means that they are set in a spin-1 representation
~\cite{Gratrex:2015hna}. We have thus to determine the transversity of the intermediate state $N$ with $\lambda=0$, with a spin equal either to 0 (scalar, pseudoscalar) or 1 (vector, axial, tensors) and $\lambda=0$. Following Ref.~\cite{Dunietz:1990cj}, it is trivial to see that $\tau(N)=0$ for spin 0. For spin 1, one can see that the $\lambda=0$ state is a superposition of states with $\tau=+1$ and $\tau=-1$, meaning that one can set $\tau(N)=1$ for spin 1. 

\begin{table}[t]
    \centering
    \begin{tabular}{c|c|c|c}
    $X$ & $\eta(N)$ & $\tau(N)$ & $\eta_X$\\    \hline
    $S$  & $1$ & 0 & $\eta$ \\
    $P$  & $-1$ & 0 & $-\eta$ \\
    $V$  & 1 & 1 & $-\eta$ \\
    $A$  & 1 & 1 & $-\eta$ \\
    $T$  & $-1$ & 1 & $\eta$  \\
    $T_t$ & 1 & 1 & $-\eta$
    \end{tabular}
    \caption{Quantum numbers and CP-parities associated with the $B\to M\ell^+\ell^-$ transversity amplitudes. We have defined $\eta=\eta(M)$. For $B_d\to K_S\ell^+\ell^-$, we have $\eta=\eta(K_S)=1$.}
    \label{tab:parities}
\end{table}

Putting these elements together yields the results collected in Table~\ref{tab:parities}, leading to
the following CP-parities associated to the different transversity amplitudes
\begin{equation}\label{eq:sumeta}
  \eta_S=  \eta_T = \eta(M) = \eta\,, \quad\quad  \eta_V=\eta_A= \eta_P=\eta_{Tt}= -\eta(M)= -\eta \ . 
\end{equation}
where we have defined $\eta=\eta(M)$. In the $B_d\to K_S\ell^+\ell^-$ case, we have $\eta(M)=\eta(K_S)=1$.

Coming back to the definition of $\bar{G}_i$, we see that the two types of angular coefficients are
related through
\begin{equation}
\widetilde{G}_i=\zeta_i \bar{G}_i\ .
\end{equation}
The number $\zeta_i$ (defined in Sec.~\ref{sec:neutralcase} to perform the identification of the kinematics between CP-conjugate decays) corresponds here
to the product of the CP-parities of the amplitudes involved in the interference term $G_i$,
for $i=0,1,2$.

\subsection{CP-averaged and CP-violating angular observables}

 We can now check the consistency of the kinematics chosen for CP-conjugate modes in Sec.~\ref{sec:neutralcase}.
 Indeed, since the decay is not flavour specific, an untagged measurement of the differential decay rate
(e.g. at LHCb, where the asymmetry production is tiny) yields:
\begin{equation}
\frac{d\Gamma(B_d\to K_S\ell\ell)+d\Gamma(\bar{B}_d\to K_S\ell\ell)}
{ds\  d\!\cos\theta_\ell}
=\sum_i [G_i+\widetilde{G}_i]  P_i(\cos\theta_\ell)
=\sum_i [G_i+\zeta_i \bar{G}_i]  P_i(\cos\theta_\ell)\ ,
\label{G+Gb}
\end{equation}
if we still neglect for the moment the effects of neutral-meson mixing.
The difference between the two decay rates (which can be measured only through flavour-tagging)
involves:
\begin{equation}
\frac{d\Gamma(B_d\to K_S\ell\ell)-d\Gamma(\bar{B}_d\to K_S\ell\ell)}
{ds\  d\!\cos\theta_\ell}
=\sum_i [G_i-\widetilde{G}_i]  P_i(\cos\theta_\ell)
=\sum_i [G_i-\zeta_i \bar{G}_i]  P_i(\cos\theta_\ell)\ .
\label{GmGb}
\end{equation}
A slightly counter-intuitive consequence of the identification of the angles between the two CP-conjugates mode is that the CP-asymmetry for $G_1$ is measured in the CP-averaged rate, and vice-versa. This situation is well known in the case of the angular distribution of other modes that are not self-tagging, such as $B_s\to\phi\ell\ell$~\cite{Bobeth:2008ij,Descotes-Genon:2015hea,Aaij:2015esa}.

We see now that the convention chosen in Eqs.~(\ref{Gamma})-(\ref{Gammabar}) for flavour-tagging modes
allows one to treat on the same footing the modes with flavour tagging and the modes with final CP-eigenstates, since the same combinations
of angular coefficients occur in both cases when one considers the CP-average or the CP-asymmetry in the decay rate.

Let us stress again that this results from a conventional identification between CP-conjugate decays in the case without mixing.
This freedom in the angular convention for CP-conjugate
flavour-specific modes
is not present in the presence of mixing where both decays result in the same final state,
which must always be described with the ``same" kinematic convention, in the sense of a convention that depends only on
the final state, without referring to the flavour of the decaying $B$ meson (see Ref.~\cite{Descotes-Genon:2015hea}). The convention chosen in Sec.~\ref{sec:neutralcase} obeys indeed this requirement and it is thus an appropriate choice even in the presence of mixing.

\subsection{Time-dependent angular distribution of $B\to K_S\ell\ell$}
\label{sec:2.4}

We can now add the effect of neutral-meson mixing. Indeed,
in the case of $B$ decays into CP-eigenstates, where the final state can be produced both by the decay of $B$ or $\bar B$
mesons, the mixing and decay processes interfere, inducing a further time dependence in physical amplitudes. These time-dependent amplitudes are given by
\begin{eqnarray}
A_X(t)&=&A_X(B(t)\to f_{CP} \ell^+\ell^-)=g_+(t) A_X + \frac{q}{p} g_-(t) \widetilde A_X\ ,\label{AXt}\\
\widetilde A_X(t)&=&A_X(\bar B(t)\to f_{CP} \ell^+\ell^-)=\frac{p}{q}g_-(t) A_X + g_+(t) \widetilde A_X\ ,
\label{AbXt}
\end{eqnarray}
where the absence of the $t$ argument denotes the amplitudes at $t=0$, i.e.~in the absence of mixing,
and we have introduced the usual time-evolution functions
\begin{eqnarray}
g_+(t)&=&e^{-imt}e^{-\Gamma t/2}\left[\cosh\frac{\Delta\Gamma t}{4}\cos\frac{\Delta m t}{2}-i\sinh\frac{\Delta\Gamma t}{4}\sin\frac{\Delta m t}{2}\right]\ ,\\
g_-(t)&=&e^{-imt}e^{-\Gamma t/2}\left[-\sinh\frac{\Delta\Gamma t}{4}\cos\frac{\Delta m t}{2}+i\cosh\frac{\Delta\Gamma t}{4}\sin\frac{\Delta m t}{2}\right]\ ,
\end{eqnarray}
with $\Delta m=M_H-M_L$ and $\Delta\Gamma=\Gamma_L-\Gamma_H$ (detailed definitions can be found in Refs.~\cite{Dunietz:2000cr,Nierste:2009wg}, which must be adapted to our choice concerning CP-conjugation in neutral meson systems: $CP|B_d\rangle=|\bar{B}_d\rangle$ and $CP|K^0\rangle=|\bar{K}^0\rangle$).

In the presence of mixing, the coefficients of the angular distribution also depend on time, as they involve the
time-dependent amplitudes given in Eqs.~(\ref{AXt}),(\ref{AbXt}). This evolution can be simplified by noting that CP violation
in $B_q-\bar B_q$ mixing is negligible for all practical purposes\footnote{The current world averages
are $|q/p|_{B_d}=1.0010\pm 0.0008$ and $|q/p|_{B_s}=1.0003\pm 0.0014$ \cite{HFLAV, PDG2019}},
and we will assume $|q/p|=1$, introducing the mixing angle $\phi$:
\begin{equation}
\frac{q}{p}=e^{i\phi}\ .
\end{equation}
This mixing angle is large in the case of the $B_d$ system but tiny for $B_s$, see Table~\ref{tab:inputs}.

The angular coefficients are obtained by replacing time-independent amplitudes with time-dependent ones:
\eq{
G_i(t) = G_i \big(A_X\to A_X(t)\big)\ ,\quad
\widetilde G_i(t) = G_i \big(A_X\to \widetilde A_X(t)\big)\ .
\label{subst}}
We consider the combinations $G_i(t) \pm \widetilde G_i(t)$ appearing in the sum and difference of time-dependent
decay rates in Eqs.~(\ref{G+Gb}),~(\ref{GmGb}). From Eqs.~(\ref{AXt}), (\ref{AbXt}) and (\ref{subst}), we get
\begin{eqnarray}\label{eq:J+Jt}
G_i(t)+\widetilde G_i(t) &=&e^{-\Gamma t}\Big[(G_i + \widetilde G_i)\cosh(y\Gamma t) - h_i \sinh(y\Gamma t)\Big]\ ,\\[2mm]
G_i(t)-\widetilde G_i(t) &=&e^{-\Gamma t}\Big[(G_i - \widetilde G_i)\cos(x\Gamma t) - s_i \sin(x\Gamma t)\Big]\ ,
\label{eq:J-Jt}
\end{eqnarray}
where $x\equiv \Delta m/\Gamma$, $y\equiv \Delta \Gamma/(2\Gamma)$, and we have defined a new set of angular
coefficients $s_i,h_i$ related to the time-dependent angular distribution.
The coefficients $G_i$, $\widetilde G_i$ can be determined from flavour-specific decays.

The expressions for $s_i$ and $h_i$ are 
\begin{eqnarray} \label{eq:s0}
s_0&=&2 {\rm Im}\left[e^{i\phi}
 \left[ \frac{4}{3} \left(1 + 2\hat{m}_\ell^2 \right) \tilde{h}_V h_V^*+  \frac{4}{3}  \beta_\ell^2 \tilde{h}_A h_A^* + 2  \beta_\ell^2 \tilde{h}_S h_S^*   +2   \tilde{h}_P h_P^*\right.\right.\\\nonumber
&&\qquad\qquad\left.\left.+ \frac{8}{3} \left( 1 + 8 \hat{m}_\ell^2  \right) \tilde{h}_{T_t} h_{T_t}^* + \frac{4}{3} \beta_\ell^2\tilde{h}_T h_T^*  
 \right]\right]
 - 16 \hat{m}_\ell \,{\rm Re}\left[e^{i\phi} \tilde{h}_V h_{T_t}^*  - e^{-i\phi}h_V\tilde{h}_{T_t}^*\right]\,,\\
 s_1&=&- 4  \beta_\ell \left(2  \hat{m}_\ell {\rm Im}\left[e^{i\phi} \tilde{h}_Vh_S^*-e^{-i\phi}h_V\tilde{h}_S^*\right]
 \right.\\\nonumber
&&\qquad\qquad\left.   +{\rm Re}\left[e^{i\phi}[2\tilde{h}_{T_t}h_S^*+\sqrt{2}\tilde{h}_Th_P^*]
                       -e^{-i\phi}[2h_{T_t}\tilde{h}_S^*+\sqrt{2}h_T\tilde{h}_P^*]\right]\right)\,,\\
s_2&=& - \frac{8  \beta_\ell^2 }{3} 
  {\rm Im}\left[e^{i\phi}\left[\tilde{h}_V h_V^*+\tilde{h}_A h_A^*
    -2\tilde{h}_Th_T^*-4\tilde{h}_{T_t} h_{T_t}^*\right]\right]\,, \label{eq:s2}
\end{eqnarray}
and
\begin{eqnarray}
h_0&=& 2{\rm Re}\left[e^{i\phi}
 \left[ \frac{4}{3} \left(1 + 2\hat{m}_\ell^2 \right) \tilde{h}_V h_V^*+  \frac{4}{3}  \beta_\ell^2 \tilde{h}_A h_A^* + 2  \beta_\ell^2 \tilde{h}_S h_S^*   +2   \tilde{h}_P h_P^*\right.\right.\\\nonumber
&&\qquad\qquad\left.\left.+ \frac{8}{3} \left( 1 + 8 \hat{m}_\ell^2  \right) \tilde{h}_{T_t} h_{T_t}^* + \frac{4}{3} \beta_\ell^2\tilde{h}_T h_T^*  
 \right]\right]
 + 16 \hat{m}_\ell \,{\rm Im}\left[e^{i\phi} \tilde{h}_V h_{T_t}^*  + e^{-i\phi}h_V\tilde{h}_{T_t}^*\right]\,,\\
 h_1&=&- 4  \beta_\ell \left(2  \hat{m}_\ell {\rm Re}\left[e^{i\phi} \tilde{h}_Vh_S^*+e^{-i\phi}h_V\tilde{h}_S^*\right]
 \right.\\\nonumber
&&\qquad\qquad\left.   -{\rm Im}\left[e^{i\phi}[2\tilde{h}_{T_t}h_S^*+\sqrt{2}\tilde{h}_Th_P^*]
                      +e^{-i\phi}[2h_{T_t}\tilde{h}_S^*+\sqrt{2}h_T\tilde{h}_P^*]\right]\right)\,,\\
h_2&=& - \frac{8  \beta_\ell^2 }{3} 
  {\rm Re}\left[e^{i\phi}\left[\tilde{h}_V h_V^*+\tilde{h}_A h_A^*
    -2\tilde{h}_Th_T^*-4\tilde{h}_{T_t} h_{T_t}^*\right]\right]\,.
\end{eqnarray}

The time-dependent angular distributions therefore contain potentially new information encoded in the new angular
observables $s_i$ and $h_i$, similarly to the ones derived in Ref.~\cite{Descotes-Genon:2015hea} for $B\to K^*\ell\ell$ and $B_s\to\phi\ell\ell$.
These observables measure the interference between $B_d$-mixing and $B\to K\ell\ell$ decay, and they contain therefore additional information compared to the angular observables presented in Sec.~\ref{sec:angobstimeindep}.

Let us stress that these observables are accessible by combining the angular distributions for $B_d(t)\to K_S\ell\ell$ and $\bar{B}_d(t)\to K_S\ell\ell$, thus requiring flavour tagging.
The coefficients $h_i$ seem very difficult to extract, since they are associated with $\sinh(y\Gamma t)$ with
$y$ vanishing at the current accuracy.
The coefficients $s_0$ and $s_2$ are associated with the CP asymmetry of angular
coefficients: $G_i-\bar G_i$, whereas $s_1$ is associated with CP-averaged angular coefficients: $G_1+\bar G_1$. 
The information on New Physics contained in the coefficients $s_i$ will be the focus of the rest of this article.

\subsection{Time-integrated observables}
\label{sec:t-int}

As discussed in Refs.~\cite{Harrison:1998yr,DescotesGenon:2011pb,Descotes-Genon:2015hea}, time integration should be performed differently in the context of
hadronic machines and $B$-factories. The time-dependent expressions in Eqs.~(\ref{eq:J+Jt}) and (\ref{eq:J-Jt})
are written in the case of tagging at a
hadronic machine, assuming that the two $b$-quarks have been produced incoherently, with $t\in [0 , \infty)$.
In the case of a coherent $B_d \bar{B}_d$ pair produced at a $B$-factory, one must replace $\exp(-\Gamma t)$
by $\exp(-\Gamma |t|)$ and integrate over $t\in (-\infty,\infty)$~\cite{Harrison:1998yr}.
Interestingly, the integrated versions of CP-violating interference terms are different in both settings,
and the measurement at hadronic machines involves an additional term compared to the $B$-factory case:
\allowdisplaybreaks{
\begin{eqnarray}
\label{eq:<J+Jt>Had}
\langle G_i + \widetilde G_i\rangle_{\rm Hadronic}
 &=& \frac{1}{\Gamma} \left[\frac{1}{1-y^2} \times(G_i+\widetilde G_i)-\frac{y}{1-y^2}\times h_i\right]\ ,\\
\label{eq:<J-Jt>Had}
\langle G_i - \widetilde G_i\rangle_{\rm Hadronic}
 &=& \frac{1}{\Gamma}\Bigg[\frac{1}{1+x^2} \times (G_i-\widetilde G_i) - \frac{x}{1+x^2} \times s_i \Bigg]\ ,\\
 \label{eq:<J+Jt>Bfac}
\langle G_i + \widetilde G_i\rangle_{\rm B-factory}
 &=& \frac{2}{\Gamma}\frac{1}{1-y^2}[G_i+\widetilde G_i]\ ,\\
  \label{eq:<J-Jt>Bfac}
 \langle  G_i - \widetilde G_i\rangle_{\rm B-factory}
 &=& \frac{2}{\Gamma}\frac{1}{1+x^2} [G_i-\widetilde G_i]\ .
\end{eqnarray}
}
 
Making contact with experimental measurements requires to consider the total time-integrated decay rate:
\begin{eqnarray}
\left\langle\frac{d(\Gamma+\bar\Gamma)}{dq^2}\right\rangle &=&  \frac{1}{\Gamma(1-y^2)} \langle{\cal I}\rangle\ ,\\[2mm]
\langle{\cal I}\rangle _{\rm Hadronic}
&=& 2(G_{0}+\bar{G}_{0}-y\,h_{0})\ ,\\[2mm]
\langle{\cal I}\rangle _{\rm B-factory} &=&2\langle{\cal I}\rangle_{\rm Hadronic}(h=0)\ ,
\end{eqnarray}
where ${\cal I}$ is the usual normalisation considered in analyses of the angular coefficients. The factor of 2 arising from the time integration in the case of $B$-factories (correcting a mistake in Ref.~\cite{Descotes-Genon:2015hea}) comes from the consideration of entangled $B\bar{B}$ pairs, leading to twice as many possibilities to observe the decay of interest compared to the hadronic case.
The normalised time-integrated angular coefficients at hadronic machines or $B$-factories are therefore:
\begin{eqnarray}
\langle \Sigma_i\rangle_{\rm Hadronic}&\equiv&
\frac{\langle G_i + \widetilde G_i\rangle_{\rm Hadronic}}{\langle d(\Gamma+\bar\Gamma)/dq^2\rangle_{\rm Hadronic}}
=\frac{(G_i+\widetilde G_i)-y\times h_i}{\langle{\cal I}\rangle_{\rm Hadronic}}\ ,\\
\langle \Sigma_i\rangle_{\rm B-factory}&\equiv&
\frac{\langle G_i + \widetilde G_i\rangle_{\rm B-factory}}{\langle d(\Gamma+\bar\Gamma)/dq^2\rangle_{\rm B-factory}}
=\langle \Sigma_i\rangle_{\rm Hadronic}(h=0)\ ,\\
\langle \Delta_i\rangle_{\rm Hadronic}&\equiv&
\frac{\langle G_i - \widetilde G_i\rangle_{\rm Hadronic}}{\langle d(\Gamma+\bar\Gamma)/dq^2\rangle_{\rm Hadronic}}
=\frac{1-y^2}{1+x^2}\times \frac{(G_i-\widetilde G_i)-x\times s_i}{\langle{\cal I}\rangle_{\rm Hadronic}}\ ,\\
\langle \Delta_i\rangle_{\rm B-factory}&\equiv&
\frac{\langle G_i - \widetilde G_i\rangle_{\rm B-factory}}{\langle d(\Gamma+\bar\Gamma)/dq^2\rangle_{\rm B-factory}}
=\langle \Delta_i\rangle_{\rm Hadronic}(h=s=0)\ .
\end{eqnarray}

We see that the interpretation of the time-integrated measurements $\langle \Sigma_i\rangle$ from
$d\Gamma(B_d\to K_S\ell\ell)+d\Gamma(\bar{B}_d\to K_S\ell\ell)$
is straightforward in terms of the angular coefficients at $t=0$. The smallness of $y$
means that $h_i$ will have only a very limited impact.
The time-integrated terms $\langle \Delta_i\rangle$  from $d\Gamma(B_d\to K_S\ell\ell)-d\Gamma(\bar{B}_d\to K_S\ell\ell)$
are subject to two different effects. On one side, they receive contributions proportional to $x$ corresponding to different combination of interference
terms (in the case of a measurement at a hadronic machine). On the other hand, they are multiplied (in all experimental set-ups) by a factor $(1-y^2)/(1+x^2)$.

We see therefore that $\av{\Sigma_i}$ contain essentially the same information as $(G_i+\widetilde G_i)$, whereas
$\av{\Delta_i}$ have a potentially richer interpretation due to the $s_i$  contribution. This contribution can be separated by comparing the time-integrated difference $d\Gamma(B\to K\ell\ell)-d\Gamma(\bar{B}\to K\ell\ell)$ in the case with mixing ($B_d\to K_S\ell\ell$) and the case without mixing ($B^+ \to K^+\ell\ell$). We have indeed (neglecting $y$)
\begin{equation}
    \langle \Delta_i\rangle_{\rm Hadronic}^{K_S}
      \equiv \frac{\langle G_i - \widetilde G_i\rangle^{K_S}_{\rm Hadronic}}{\langle d(\Gamma+\bar\Gamma)/dq^2\rangle^{K_S}_{\rm Hadronic}}
      =\frac{(G_i-\widetilde G_i)-x s_i}{2(1+x^2)(G_0+\bar{G}_0)}
\end{equation}
leading to
\begin{equation}\label{eq:Delta012}
    \langle \Delta_i\rangle_{\rm Hadronic}^{K_S}
   =\frac{1}{1+x^2} \langle \Delta_i\rangle^{K^\pm}
      -\frac{x}{1+x^2} \sigma_i\, \qquad \sigma_i=\frac{s_i}{2
      \Gamma_\ell}\, \qquad i=0,1,2\,.
\end{equation}

We have~\cite{PDG2019}
\begin{equation}
 \frac{1}{1+x^2}=0.6284(24)\,,
 \qquad  \frac{x}{1+x^2}=0.4832(6)\,,
\end{equation}
showing that there is a good sensitivity 
to $\sigma_i$
using time-integrated observables.
We will show that in the SM and in any NP extension with SM operators and chirally flipped operators, we obtain a very precise prediction for the $\sigma_i$. Therefore, also the relation between $ \langle \Delta_i\rangle_{\rm Hadronic}^{K_S}$ and $\langle\Delta_i\rangle^{K^\pm}$ can be predicted with high precision which is a very powerful and generic test of the structure of the operators contributing (real, no scalars, no tensors).

Let us add that these time-integrated observables still require a flavour tagging to separate the decays originating from a $B_d$ meson from the ones starting from a $\bar{B}_d$-meson, i.e.~$d\Gamma(B_d\to K_S\ell\ell)$ and $d\Gamma(\bar{B}_d\to K_S\ell\ell)$. Therefore, this approach enables one to bypass the study of the time dependence, but it still requires initial-state flavour tagging (with the associated effective loss of statistical power).

\subsection{Extension to other $B_d$ and $B_s$ decays into light spin-0 mesons
%}$B_s\to f_0\ell\ell$ and $B_d\to K_L\ell\ell$
}\label{sec:extensions}

Our analysis applies to any $B\to P\ell\ell$ decay where the initial neutral meson mixes with its antimeson and the final meson $P$ is a (scalar or pseudoscalar) spin-0 CP eigenstate. 

Another mode that could be considered is $B_d\to K_L\ell\ell$. The opposite intrinsic parity of $K_L$ with respect to $K_S$ means that $\widetilde{G}_i=-\zeta_i\bar{G}_i$ and
\begin{equation}
\frac{d\Gamma(B_d\to K_L\ell\ell)\pm d\Gamma(\bar{B}_d\to K_L\ell\ell)}
{ds\  d\!\cos\theta_\ell}
=\sum_i [G_i\pm \widetilde{G}_i]  P_i(\cos\theta_\ell)
=\sum_i [G_i\mp \zeta_i \bar{G}_i]  P_i(\cos\theta_\ell)\ ,
\end{equation}
where the $G_i$ ($\bar{G}_i$) are defined in terms of $h_X(B_d\to K_L)$ ($\bar{h}_X(\bar{B}_d\to K_L)$). In the absence of CP violation, we would have $G_i=-\bar{G}_i$ due to the different normalisation for $h_X(B_d\to K_L)$ and $\bar{h}_X(\bar{B}_d\to K_L)$. The discussion concerning time-dependent observables is unchanged. We see that the most promising observables $s_{0,1,2}$ can still be accessed through the difference $d\Gamma(B_d\to K_L\ell\ell)- d\Gamma(\bar{B}_d\to K_L\ell\ell)$. However, due to the additional experimental difficulties related to the identification of the $K_L$ meson, we will focus on the $K_S$ case in the following.

One can also consider $B_s$ decays.
The mixing parameters are different 
from the $B_d$ case since $x$ is much larger (whereas $y$ is small but not vanishing) and the mixing angle $2\beta_s$ is very small~\cite{PDG2019}.
In this case, the coefficients $h_i$ are difficult to extract, since they are associated with $\sinh(y\Gamma t)$ with
$y$ small, but at least not vanishing, meaning that an extraction of $h_i$ is possible in this case. The $s_i$ coefficients are certainly easier to access, but they are different from zero only if there are large NP phases or large tensor contributions, which are not needed in the current global fits to $b\to s\ell\ell$ transitions (see for instance Ref.~\cite{Alguero:2019ptt}).

One may consider $B_s\to f_0(980)\ell\ell$ with
$\eta(f_0)=\eta(K_S)=1$ and $B_s\to \eta(')\ell\ell$ with $\eta(\eta('))=-1$.
The determination of the form factors is quite challenging in all three cases, as their exact nature and mixing with other states are not known precisely yet. This translates into a significant spread of results for the form factors. We will thus focus in the following on $B_d\to K_S\mu\mu$ which is better understood from the theory point of view, but we will give a few results for the $B_s$ to $f_0, \eta, \eta'$ decays in Appendix~\ref{app:f0etaetaprime}.

\section{New observables in $B_d\to K_S\mu\mu$ as probes of new physics} \label{sec:observables}

\subsection{Real NP contributions to SM and chirally flipped Wilson coefficients}\label{sec:realNPSM}

We will now consider the normalised observables 
\begin{equation}\label{eq:sigmarho}
    \sigma_i=\frac{s_i}{2(G_0 + \bar{G}_0)}=\frac{s_i}{2\Gamma_\ell}\,, \qquad
    \rho_i=\frac{s_i}{2(G_i + \bar{G}_i)}\,,
    \qquad i=0,1,2\,,
\end{equation}
where the normalisation comes from the CP-averaged decay rate
$\Gamma_\ell=G_0+\bar{G}_0$. We set $y=0$ and we will neglect the tiny weak phase in $V_{tb}V_{ts}^*$ in the following.

We start by considering scenarios where NP enters only as real shifts to the Wilson coefficients for SM and chirally flipped operators (${\cal C}_{7,7',9,9',10,10'}$). This case includes naturally the SM, but it also covers many NP scenarios currently favoured by global fits to $b\to s\ell\ell$ data~\cite{Alguero:2019ptt}.
In this case, we have only contributions from the amplitudes $h_V$, $h_A$ and $h_P$. Neglecting the (tiny) weak phase in $V_{tb}V_{ts}^*$, gives $\tilde{h}_{V,A,P}=-h_{V,A,P}$ and $G_i=\bar{G}_i$ ($i=0,1,2$). 

The $s_i$ observables then become very simple, leading to~\footnote{If we neglect the lepton mass, we have also $\sigma_2=-\sigma_0$.}
\begin{equation}\label{eq:realNPrelations}
    \rho_0=\rho_2=\sigma_0 = -\frac{\sin\phi}{2} \ , \quad\quad \sigma_1=0 \,, \qquad \phi=-2\beta\,.
 \end{equation}
We stress that these relations neither depend on a specific choice of form factors nor on assumptions made on charm-loop contributions. The only hypothesis required is that NP enters as real contributions to the SM Wilson coefficients. Therefore, a measurement of these observables would constitute a significant cross-check of the NP scenarios currently favoured by global fits to $b\to s\ell\ell$ data~\cite{Alguero:2019ptt}. Moreover, the only parameter with a non-trivial but very small $q^2$-dependence at the kinematic endpoints is $\sigma_2$, such that the relations Eq.~(\ref{eq:realNPrelations}) can be checked by integrating over any kinematic $q^2$ range. On the other hand, a deviation from these values would constitute a very simple and powerful test of the presence of scalar/tensor operators or that of  CP-violating NP phases. We discuss these two cases next.

\subsection{Real NP contributions including scalar and tensor operators}
\label{sec:NPscalartensor}

Considering still real NP contributions, but adding possible scalar and tensor contributions, changes the above situation.
The expressions for $G_i$ and $s_i$ can be reduced in the following way (neglecting $m_\ell$ effects for simplicity): 
\begin{eqnarray}
\label{eq:gi_real_NP_massless}
G_0 (= \bar{G}_{0}) &\simeq & \frac{4}{3} \abs{\HVK}^2 +  \frac{4}{3} \abs{\HAK}^2  + 2  \abs{\HSK}^2   +2   \abs{\HPK}^2 + \frac{8}{3}  \abs{\HTenstK}^2 + \frac{4}{3} \abs{\HTensK}^2   \; , \nonumber \\
G_1 (= \bar{G}_{1}) &\simeq &  0 \; , \nonumber \\
G_2 (=\bar{G}_{2}) &\simeq & - \frac{4}{3} \left( \abs{\HVK}^2 + \abs{\HAK}^2 - 2 \abs{\HTensK}^2 - 4 \abs{\HTenstK}^2\right) \nonumber \\
&\simeq & -G_0 +  2  \abs{\HSK}^2   +2   \abs{\HPK}^2 + 8 \abs{\HTenstK}^2 + 4 \abs{\HTensK}^2   \,,
\end{eqnarray}
leading to
\begin{eqnarray}
\label{eq:si_real_NP_massless}
s_0 &\simeq & -2 \sin \phi \left(G_0 - 4  \abs{\HSK}^2 - \frac{16}{3} \abs{\HTenstK}^2\right)   \; , \nonumber \\
s_1 &\simeq &  8 \sin\phi \left(-2{\rm Im} [\HTenstK]h_S^*+\sqrt{2}{\rm Im}[\HTensK] h_P^*\right) \; , \nonumber \\
s_2 &\simeq & -2 \sin \phi \left(G_2 - \frac{32}{3} \abs{\HTenstK}^2\right)   \; ,
\end{eqnarray}
Observing how these observables depend on the different scalar, pseudoscalar and tensor contribution allows us to define new observables separating these contributions:
\begin{align}\label{eq:RS}
R_S\equiv \frac{2}{\sin\phi}(-\sigma_2+2\sigma_0)-F_H^\ell+3
 &\simeq  16\frac{|\bar{h}_S|^2}{\Gamma_\ell}\,,\\
R_{T_t}\equiv \frac{2}{\sin\phi} \sigma_2+F_H^\ell-1&\simeq 
 \frac{64}{3}\frac{|\bar{h}_{T_t}|^2}{\Gamma_\ell} \,.\label{eq:RTt}
\end{align}
These observables could be obtained from a joint study of the charged and neutral $B\to K\ell\ell$ decays.
Neglecting $m_\ell$-suppressed contributions,
$R_S$ and $R_{T_t}$ allow for a neat separation of the scalar and tensor contributions, contrary to the CP-averaged observables, and in this limit, these two observables must be positive in the absence of NP complex phases.

The combination 
\begin{equation}\label{eq:RW}
\begin{split}
R_W&\equiv R_S+3R_{T_t}=\frac{4}{\sin\phi} \left(\sigma_0+\sigma_2\right)+2F_H^\ell\\&=
\frac{2}{\sin\phi \Gamma_\ell}
 [s_0+s_2+\sin\phi(G_0+\bar{G_0}+G_2+\bar{G_2})]
\simeq  
\frac{16}{\Gamma_\ell}[|\bar{h}_S|^2+4|\bar{h}_{T_t}|^2]
\end{split}
\end{equation}
is also interesting. It vanishes exactly in the limit where $m_\ell=0$ and ${\cal C}_S={\cal C}_P={\cal C}_T={\cal C}_{T_t}=0$, no matter what the values (real or complex) for ${\cal C}_{7,7',9,9',10,10'}$. Indeed, in this limit, $G_0=-G_2$, $\bar{G}_0=-\bar{G}_2$ and $s_0=-s_2$, as can be checked explicitly from Eqs.~(\ref{eq:gi}), (\ref{eq:s0}) and (\ref{eq:s2}). One can thus expect that the deviations of $R_W$ from zero should be rather sensitive to the presence of scalar and tensor contributions.

When accounting for $m_\ell$ and the tiny imaginary part of $V_{ts}$, 
these new observables do not vanish any more in the SM. We give their SM values in Table~\ref{tab:3scen} over the bin in $q^2$ from 1 to 6 GeV$^2$ using the inputs specified in Table~\ref{tab:inputs}. SM predictions at different values of $q^2$ or specific bins can easily be obtained using the above equations. We give in Appendix~\ref{tab:models} the results using alternative models for the charm-loop contribution, showing a very good stability of our results with respect to the change of model, covered by our theoretical uncertainties~\footnote{Neglected doubly Cabibbo suppressed contributions with relative size of $O(\lambda^2)\simeq 4\%$ are not included in our estimate of the uncertainties, but they do not affect our conclusions concerning the capacity of theses observables to discriminate NP scenarios.}

The sensitivity to NP scalar and tensor contributions of these observables is
\begin{align}\label{eq:rsrt}
    R_S &    = 0.028 | {\mathcal C}_S + {\mathcal C}_{S}|^2 \ , \nonumber \\
    R_{T_t} & = 0.019|{\mathcal C}_{T} + {\mathcal C}_{T'}|^2  \ , \nonumber \\
     R_{W} &=  0.028 |{\mathcal C}_S + {\mathcal C}_{S'}|^2 + 0.056 |{\mathcal C}_T + {\mathcal C}_{T'}|^2 \ .
\end{align}
Currently, the bounds on scalar contributions are quite loose.
Ref.~\cite{Arbey:2019duh} suggest $|{\cal C}_{S\mu}|<0.1$
    and $0<{\cal C}_{S'\mu}<0.2$  obtained for NP models containing also SM-like and chirally flipped operators with real Wilson coefficients. We are not aware of studies giving bounds on tensor operators, probably due to the fact there are currently no indication of a need for such contributions in global fits. 
    
    In order to illustrate the effect of new scalar or tensor contributions, we consider two NP scenarios with ${\cal C}_{S}=0.2$ and ${\cal C}_{T}=0.2$, respectively. Although $R_S$ and $R_{T_t}$ are in principle sensitive to scalar and tensor operators, we see that the changes are rather small, as is expected from \eqref{eq:rsrt}. The situation is different for $R_W$, which is constructed such that it vanishes exactly in the absence of scalar and tensor corrections in the limit $m_\ell=0$. The SM value is different from zero due to $m_\ell$-suppressed corrections, but its value is known very precisely, and it deviates from this value when scalar and/or tensor contributions are present.
    
    Given the accuracy of the theory predictions,
    it seems thus possible to gain information on scalar and tensor contributions from $R_S$, $R_{T_t}$ and $R_{W}$ if they are measured  precisely, in complement with the information provided by $F_H^\ell$.

\subsection{Complex NP contributions}
\label{sec:NPcomplex}

The equalities in Eq.~(\ref{eq:realNPrelations}) do not hold in the presence of complex NP contributions. In principle, these contributions can be constrained by measuring $\Gamma^\ell$ and $A_{FB}^\ell$, but their effect in those observables is suppressed by $m_\ell$. Besides, such NP effects would show up in the direct CP-asymmetries $A_0$ and $A_2$ but due to the interferences between weak and strong phases in those observables the interpretation is less clear. Moreover, as can be seen in Appendix~\ref{app:predictions}, hadronic uncertainties are significant for these observables compared to their sensitivity to NP, so that it is difficult to interpret a deviation from the SM expectations~\footnote{It was recently proposed to consider the CP asymmetries in the vincinity of charmonium resonances to enhance their values~\cite{Becirevic:2020ssj}}.

Our new observables $s_i$ correspond to an interference between mixing and decay, and thus are sensitive to NP phases coming from all the amplitudes $h_X$ and all Wilson coefficients. As an illustration of the added power of these observables, we can use 
Ref.~\cite{Biswas:2020uaq}   where the following scenarios obtain a good description of the data with the following
best-fit points:
\begin{eqnarray}\label{eq:3scen}
{\rm Scenario\ 1}&:&{\cal C}_{9\mu}^{\rm NP} = -1.12 + i 1.00 \ , \nonumber \\
{\rm Scenario\ 2}&:&
{\cal C}_{9\mu}^{\rm NP} = -1.14-i0.22\,,\qquad
{\cal C}_{9'\mu}^{\rm NP} =
0.40-i0.38 \ , \nonumber \\
{\rm Scenario\ 3}&:&
{\cal C}_{9\mu}^{\rm NP} =-1.13-i0.12\,,\qquad
{\cal C}_{9'\mu} =0.52-i1.80\,,\qquad
{\cal C}_{10\mu}^{\rm NP} =0.41+i0.13 \ ,
\end{eqnarray}
In these scenarios, we have still $\sigma_1=\rho_1=0$, but the situation is rather different for the cases $\sigma_{0,2}$. The resulting predictions integrating over the bin in $q^2$ from 1 to 6 GeV$^2$ are given in Table~\ref{tab:3scen} using in addition the inputs in Table~\ref{tab:inputs}, i.e.~including $m_\ell= m_\mu$ and the imaginary part of $V_{ts}$. In addition, we give the values for $R_S, R_{T_t}$ and $R_W$ for the three NP scenarios in \eqref{eq:3scen}. We give in Appendix~\ref{app:predictions} the results using alternative models for the charm-loop contribution, showing a very good stability of our results with respect to the change of model. Moreover, our uncertainties cover the small changes in the central values when we consider different models for the $c\bar{c}$ contributions. The values in Table~\ref{tab:3scen} serve as an illustration of the sensitivity of our observables to the three new physics scenarios: using our expressions, values for different $q^2$ ranges can be easily obtained. We note that the uncertainties in Table~\ref{tab:3scen} for $\sigma_0$ and $\sigma_2$ are fully dominated by the uncertainty on $2\beta$.  

We observe that although $\sigma_0$ and $\sigma_2$ are sensitive to the three NP scenarios, in fact $R_S$ and $R_{T_t}$ are even more sensitive. The deviations of the latter two observables from their SM expectation values allows for a distinction between the three different NP scenarios, even once hadronic uncertainties are taken into account. $R_S$ and $R_{T_t}$ are thus interesting probes for these new weak phases, whereas $R_W$ is still very small in these scenarios (it would vanish in the limit where $m_\ell$ vanishes).

We emphasize that the above scenarios only serve as a benchmark to indicate the sensitivity of the observables to new phases. Once experimental measurements of these observables are available, performing a more sophisticated NP analysis, including scalar, tensor and complex phases would be interesting.

\begin{table}[t]
\begin{center}
\begin{tabular}{c|c|c|c|c|c|c}
Observable & SM &  Scen. 1 & Scen. 2 &  Scen. 3 &$C_S=0.2$&$C_T=0.2$\\
\hline
$\sigma_0 $ & $0.368(5)$ & $0.273(6)$ & $0.402(5)$ & $0.43(1)$ & $0.368(5)$ & $0.368(5)$\\
$\sigma_2$ & $-0.359(5)$ & $-0.266(6)$&$-0.392(4)$&$-0.415(9)$& $-0.359(5)$& $-0.357(5)$
\\
$R_S $ & $-0.107(4)$ & $0.69(2)$ & $-0.39(2)$ & $-0.59(9)$ & $-0.105(4)$& $-0.107(4)$\\
$R_{T_t}$ & $0.035(1)$& $-0.225(8)$ &$0.128(7)$ & $0.19(3)$& $0.035(1)$& $0.036(1)$ \\
$R_{W}\times10^2$ & $-0.179(8)$& $1.09(4)$&$-0.63(4)$&$-1.0(1)$&$-0.01(1)$&$0.04(3)$ 
\end{tabular}
\caption{Values of the observables in the SM, for the three different scenarios with new complex Wilson coefficients defined in~\eqref{eq:3scen} and for the scenarios with ${\cal C}_{S(T)}=0.2$. All quoted values are for $B_d\to K_S\mu\mu$ and are binned in $q^2$ over $[1,6]$ GeV$^2$. The inputs are taken from Table~\ref{tab:inputs} (neglected doubly Cabibbo-suppressed contributions are not included in our error estimates). A more comprehensive list of results is given in Table~\ref{tab:models}. Values for other fixed $q^2$ values of specific bins can be easily obtained from our expressions.}\label{tab:3scen}
\end{center}
\end{table}

\subsection{New physics benchmarking from $B_d\to K_S\mu\mu$}
We have seen that a time-dependent angular analysis of $B_d\to K_S\mu\mu$ leads to 6 new observables, measuring CP violation in the interference between decay and mixing. 
They can be obtained from:
\begin{eqnarray}
&&\frac{d\Gamma(B_d(t)\to K_S\ell\ell)-d\Gamma(\bar{B}_d(t)\to K_S\ell\ell)}
{ds\  d\!\cos\theta_\ell }\nonumber\\
&&\qquad=[G_0-\widetilde{G}_0](t)
  +[G_1-\widetilde{G}_1](t)\,\cos\theta_\ell
  +[G_2-\widetilde{G}_2](t)\,\frac{1}{2}(3\cos^2\theta_\ell-1)
\end{eqnarray}
with the time dependence described in Eq.~(\ref{eq:J-Jt}):
\begin{equation}
G_i(t)-\widetilde G_i(t) = e^{-\Gamma t}\Big[(G_i - \widetilde G_i)\cos(x\Gamma t) - s_i \sin(x\Gamma t)\Big]\ ,
\end{equation}
We also showed that time-integrated angular observables could also provide a good sensitivity on the coefficients $s_{0,1,2}$, by comparing neutral and charge modes at hadronic machines. These observables can be predicted accurately. Depending on the NP scenario, their theoretical predictions have little to no sensitivity to the specific choices for the form factors or the charm-loop contributions.

These three observables can be combined with the usual angular observables for $B\to K\ell\ell$ to obtain the observables $\sigma_{0,1,2}$,  $\rho_{2}$ and $R_{S,T_t,W}$ defined in Eqs.~(\ref{eq:sigmarho}), (\ref{eq:RS}), (\ref{eq:RTt}) and (\ref{eq:RW}), respectively. These quantities can be computed very precisely theoretically (see Table~\ref{tab:3scen}). 
If measured precisely, these observables provide powerful probes for New Physics scenarios:
\begin{itemize}
    \item Do $\sigma_0,\sigma_1,\rho_2$ obey the simple relations in Eq.~(\ref{eq:realNPrelations}), directly related to $B_d$-$\bar{B}_d$ mixing? \\
    If yes, NP enters only the SM and chirally flipped operators
    $\op_{7('),9('),10(')}$ with real contributions, in agreement with the NP scenarios currently favoured by global fits to $b\to s\ell\ell$ data.
    \item Do $\sigma_0$, $\sigma_2$, $R_S$ and/or $R_{T_t}$ deviate from their SM expectations?\\ If yes, it means that NP enters with imaginary contributions, odd under CP-conjugation.
    \item Does $R_W$ deviate from its SM expectation, but are $\sigma_0$, $\sigma_2$, $R_S$ and $R_{T_t}$ close to the SM?\\
    If yes, it means that NP enters through scalar and tensor contributions. Complementary information is then obtained through $F_H^\ell$.
\end{itemize}
We thus see that the time-dependent angular analysis of $B_d\to K_S\mu\mu$ yields interesting observables for the discrimination among NP scenarios, if they can be measured with a sufficient precision~\footnote{We have focused on $B_d\to K_S\mu\mu$ due to the current hints of New Physics in $b\to s\mu\mu$ transitions, but similar measurements in the electron sector would also be of interest to probe lepton flavour universality.}. This could be achieved in particular at LHCb, where the time-integrated observables have already been measured~\cite{Aaij:2014tfa} and Belle II, following the measurements of branching ratios already performed at Belle~\cite{Abdesselam:2019lab}. Determining the potential of these two experiments for theses measurements is an interesting question which we leave for future work.

As already been discussed in Sec.~\ref{sec:extensions}, a similar approach could be used for $B_s$ decays such as $B_s\to f_0\mu\mu$, $B_s\to\eta\mu\mu$, $B_s\to \eta'\mu\mu$. The theoretical determination of the relevant form factor becomes complicated due to the debated structure of these mesons, but we discuss a few results regarding these decays in Appendix~\ref{app:f0etaetaprime}.

\section{Conclusions} \label{sec:conclusions}

The recent measurements of $b\to s\mu\mu$ transitions
led to tantalizing hints of New Physics. It is thus particularly important to probe these transitions with a higher experimental and theoretical accuracy, but also to provide new modes and observables constraining NP scenarios in different ways.

One approach consists in using neutral-meson mixing and time-dependent analysis in order to define new observables for $B_d$ and $B_s$ decays. This was applied to $B_d\to K^*\mu\mu$ and $B_s\to \phi\mu\mu$ in Ref.~\cite{Descotes-Genon:2015hea}. In this article, we considered the same idea in the simpler case of $B_d\to K_S\mu\mu$. The charged mode $B^\pm \to K^\pm \mu\mu$ has a much simpler angular structure, with only three observables which provide interesting but limited
constraints on scalar and tensor contributions. The hadronic inputs (form factors and charm-loop contributions) are also much more simple to handle and analyse. We discussed the benefits of a time-dependent angular analysis of this mode.

After recalling the formalism in the absence of mixing (charged case), we turned to the neutral case. It required a careful definition of the kinematics of the mode to connect CP-conjugate decays that are now related through $B_d$ mixing into $\bar{B}_d$. A time-dependent angular analysis leads to 6 new observables, measuring CP violation in the interference between decay and mixing. Three of these observables, denoted  $s_{0,1,2}$, seem rather promising, and they can also be obtained from time-integrated angular observables  by comparing neutral and charge modes at hadronic machines if initial-state flavour tagging is available.
These 3 observables $s_{0,1,2}$ have simple expressions in terms of the transversity amplitudes given in Eqs.~(\ref{eq:s0})-(\ref{eq:s2}). They can be combined  with the usual angular observables for $B\to K\ell\ell$ to obtain the observables $\sigma_{0,1,2}$,  $\rho_{2}$ and $R_{S,T_t,W}$ defined in Eqs.~(\ref{eq:sigmarho}), (\ref{eq:RS}), (\ref{eq:RTt}) and (\ref{eq:RW}).

Very interestingly, we showed that $\sigma_0$ and $\rho_2$
are very precisely known in the Standard Model and in New Physics scenarios with real contributions to SM and chirally-flipped operators. They depend only on the $B_d$-mixing angle, i.e.~the CKM angle $\beta$, see Eq.~(\ref{eq:realNPrelations}), and they are valid for any value of the dilepton invariant mass $q^2$.
We stress that these predictions are very robust, as they  hold no matter what the assumptions on form factors and charm-loop contributions are. Therefore they constitute very powerful probes of the NP scenarios currently favoured by the global fits to $b\to s\ell\ell$ data.

We have then investigated two NP cases where these predictions are modified. We showed that $R_S$, $R_T$ and $R_W$ 
can probe other NP scenarios, namely scalar and tensor operators (with real contributions) and complex NP contributions entering with a CP-odd ``weak'' phase.
The sensitivity of these observables  to each scenario is different, and the theoretical uncertainties attached to the theoretical predictions are small, which allows us to provide a benchmark of NP scenarios hinging on the measurements of $s_{0,1,2}$. We also briefly discussed similar $B_s$ decay modes such as $B_s\to f_0\mu\mu$, $B_s\to \eta\mu\mu$ and $B_s\to \eta'\mu\mu$.

In conclusion, the simplicity of the underlying $B\to K\mu\mu$ decay has allowed us to provide a detail analysis
of the flavour-tagged time-dependent analysis of $B_d\to K_S\mu\mu$. These new observables provide powerful cross checks of the various NP hypotheses. They may also contribute to global fits to $b\to s\ell\ell$ in a useful way, providing constraints of a different type on the short-distance physics encoded in Wilson coefficients. Due to the potential of $b\to s\ell\ell$ transitions  to open windows of the physics beyond the Standard Model, it is clear that the determination
and measurement of new observables will
 remain a topic of intense discussion both experimentally and theoretically in the coming years, and we hope that $B_d\to K_S\mu\mu$ (and similar $B_s$ decays) will contribute to the field in a useful manner.

\section*{Acknowledgments}
We would like to thank Christoph Langenbruch and Karim Trabelsi for discussions and comments.
We thank Bla\v{z}enka Meli\'c for sending us updated results for Ref.~\cite{Duplancic:2015zna}. The work of KKV is supported by the DFG Sonderforschungsbereich/Transregio 110 ``Symmetries and the Emergence of Structure in QCD''.

\appendix

\section{Predictions for $B_d\to K_S\mu\mu$ observables in SM and NP scenarios}\label{app:predictions}

\begin{table}[ht]
    \centering
    \begin{adjustbox}{width=1.\textwidth,center=\textwidth}
    \begin{tabular}{|c|c|c|c|c|c|c|c|}
  \cline{3-8}
\multicolumn{2}{c|}{$\quad$}   & SM & Scenario 1 & Scenario 2 & Scenario 3 & $\Cc{S}=0.2$& $\Cc{T}=0.2$\\ \hline
\multirow{3}{*}{$Br\times 10^8$}& O.E. & $ 8.4\pm 1.5$ & $ 6.8\pm 1.2$ & $ 7.2\pm 1.2$ & $ 7.4\pm 1.3$ & $ 8.4\pm 1.5$ & $ 8.4\pm 1.5$ \\ \cline{2-8}
  & LCSR & $ 7.9\pm 1.3$ & $ 6.5\pm 1.0$ & $ 6.9\pm 1.1$ & $ 7.0\pm 1.1$ & $ 7.9\pm 1.3$ & $ 8.0\pm 1.3$ \\ \cline{2-8}
  & PQCD & $ 8.6\pm 1.4$ & $ 7.0\pm 1.1$ & $ 7.4\pm 1.2$ & $ 7.6\pm 1.2$ & $ 8.6\pm 1.4$ & $ 8.7\pm 1.4$ \\ \hline \hline
\multirow{3}{*}{$F^{\ell}_H\times 10^2$}& O.E. & $ 2.48\pm 0.04$ & $ 2.50\pm 0.04$ & $ 2.50\pm 0.04$ & $ 2.48\pm 0.03$ & $ 2.52\pm 0.04$ & $ 3.05\pm 0.05$ \\ \cline{2-8}
  & LCSR & $ 2.49\pm 0.04$ & $ 2.51\pm 0.04$ & $ 2.50\pm 0.04$ & $ 2.48\pm 0.03$ & $ 2.53\pm 0.04$ & $ 3.05\pm 0.05$ \\ \cline{2-8}
  & PQCD & $ 2.49\pm 0.03$ & $ 2.51\pm 0.04$ & $ 2.50\pm 0.04$ & $ 2.48\pm 0.03$ & $ 2.53\pm 0.03$ & $ 3.05\pm 0.05$ \\ \hline \hline
\multirow{3}{*}{$A_0\times 10^2$}& O.E. & 0 & $ 0.0\pm 2.3$ & $ 0.0\pm 1.3$ & $ 0.\pm 4.$ & 0 & 0 \\ \cline{2-8}
  & LCSR & 0 & $ 3.9\pm 2.1$ & $ -2.2\pm 1.2$ & $ -7.\pm 4.$ & 0 & 0 \\ \cline{2-8}
  & PQCD & 0 & $ -1.285\pm 0.005$ & $ 0.729\pm 0.003$ & $ 2.27\pm 0.01$ & 0 & 0 \\ \hline \hline
\multirow{3}{*}{$\sigma _0\times 10$}& O.E. & $ 3.68\pm 0.05$ & $ 2.73\pm 0.06$ & $ 4.02\pm 0.05$ & $ 4.3\pm 0.1$ & $ 3.68\pm 0.05$ & $ 3.68\pm 0.05$ \\ \cline{2-8}
  & LCSR & $ 3.68\pm 0.05$ & $ 2.77\pm 0.06$ & $ 3.99\pm 0.04$ & $ 4.15\pm 0.04$ & $ 3.68\pm 0.05$ & $ 3.68\pm 0.05$ \\ \cline{2-8}
  & PQCD & $ 3.68\pm 0.05$ & $ 2.73\pm 0.06$ & $ 4.03\pm 0.04$ & $ 4.29\pm 0.01$ & $ 3.68\pm 0.05$ & $ 3.68\pm 0.05$ \\ \hline \hline
\multirow{3}{*}{$\sigma _2\times 10$}& O.E. & $ -3.59\pm 0.05$ & $ -2.66\pm 0.06$ & $ -3.92\pm 0.04$ & $ -4.15\pm 0.09$ & $ -3.59\pm 0.05$ & $ -3.57\pm 0.05$ \\ \cline{2-8}
  & LCSR & $ -3.59\pm 0.05$ & $ -2.69\pm 0.05$ & $ -3.89\pm 0.04$ & $ -4.05\pm 0.04$ & $ -3.59\pm 0.05$ & $ -3.57\pm 0.05$ \\ \cline{2-8}
  & PQCD & $ -3.59\pm 0.05$ & $ -2.66\pm 0.05$ & $ -3.92\pm 0.04$ & $ -4.18\pm 0.01$ & $ -3.59\pm 0.05$ & $ -3.57\pm 0.05$ \\ \hline \hline
\multirow{3}{*}{$R_S\times 10$}& O.E. & $ -1.07\pm 0.04$ & $ 6.9\pm 0.2$ & $ -3.9\pm 0.2$ & $ -5.9\pm 0.9$ & $ -1.05\pm 0.04$ & $ -1.07\pm 0.04$ \\ \cline{2-8}
  & LCSR & $ -1.07\pm 0.04$ & $ 6.6\pm 0.2$ & $ -3.7\pm 0.2$ & $ -5.0\pm 0.5$ & $ -1.05\pm 0.04$ & $ -1.07\pm 0.04$ \\ \cline{2-8}
  & PQCD & $ -1.07\pm 0.04$ & $ 6.9\pm 0.1$ & $ -4.0\pm 0.1$ & $ -6.2\pm 0.4$ & $ -1.05\pm 0.04$ & $ -1.07\pm 0.04$ \\ \hline \hline
\multirow{3}{*}{$R_{T_t}\times 10$}& O.E. & $ 0.35\pm 0.01$ & $ -2.25\pm 0.08$ & $ 1.28\pm 0.07$ & $ 1.9\pm 0.3$ & $ 0.35\pm 0.01$ & $ 0.36\pm 0.01$ \\ \cline{2-8}
  & LCSR & $ 0.35\pm 0.01$ & $ -2.16\pm 0.06$ & $ 1.21\pm 0.05$ & $ 1.7\pm 0.2$ & $ 0.35\pm 0.01$ & $ 0.36\pm 0.01$ \\ \cline{2-8}
  & PQCD & $ 0.35\pm 0.01$ & $ -2.27\pm 0.05$ & $ 1.30\pm 0.05$ & $ 2.0\pm 0.1$ & $ 0.35\pm 0.01$ & $ 0.36\pm 0.01$ \\ \hline \hline
\multirow{3}{*}{$R_W\times 10^2$}& O.E. & $ -0.179\pm 0.008$ & $ 1.09\pm 0.04$ & $ -0.63\pm 0.04$ & $ -1.0\pm 0.1$ & $ -0.01\pm 0.01$ & $ 0.04\pm 0.03$ \\ \cline{2-8}
  & LCSR & $ -0.179\pm 0.008$ & $ 1.05\pm 0.03$ & $ -0.60\pm 0.03$ & $ -0.83\pm 0.08$ & $ 0.000\pm 0.009$ & $ 0.05\pm 0.03$ \\ \cline{2-8}
  & PQCD & $ -0.179\pm 0.008$ & $ 1.10\pm 0.02$ & $ -0.65\pm 0.02$ & $ -1.01\pm 0.07$ & $ -0.013\pm 0.008$ & $ 0.03\pm 0.03$ \\ \hline
\end{tabular}
    
   \end{adjustbox}
    \caption{$B_d\to K_S\mu\mu$ observables integrated from $1-6\ {\rm GeV}^2$ for different parametrisations of the $c\bar{c}$ model contribution associated with $\Cc9$ (Our Estimate (O.E.), Light-Cone Sum Rules (LCSR), Perturbative QCD (PQCD)).
    The results are given in the SM case, for several NP scenarios with weak phases (Scenarios 1,2,3) and with contribution to scalar and tensor 
    contributions (${\cal C}_S,{\cal C}_T$). Neglected doubly Cabibbo-suppressed contributions are not included in our error estimates.
    }
    \label{tab:models}
\end{table}

We give in Table~\ref{tab:models} our predictions
for the various observables of interest, integrated over the bin [1,6] GeV$^2$ for the dimuon invariant mass. We compute their values
within the SM and several different scenarios of NP, described in Secs.~\ref{sec:NPscalartensor} and
\ref{sec:NPcomplex}. We neglect doubly Cabibbo-suppressed contributions with relative size of $O(\lambda^2)\simeq 4\%$, which are not included in our error estimates.

We take into account the various sources of uncertainties (CKM, form factors, charm-loop contributions) and combine them in quadrature. We illustrate the impact of the model used for charm-loop contributions by considering three models described in Sec.~\ref{sec:hadronicinputs}:
\begin{itemize}
    \item O.E.: Our estimate, corresponding to Eq.~(\ref{eq:C9ccmodel}),
    \item LCSR: A range inspired by the Light-Cone Sum Rule Estimate of Ref.~\cite{Khodjamirian:2012rm}, given in Eq.~(\ref{eq:C9ccLCSR}),
    \item PQCD: The short-distance charm-loop contribution obtained from perturbative QCD,
    illustrated in Eq.~(\ref{eq:C9ccpert}),
    without attaching any uncertainty to the result.
\end{itemize}
As can be seen, our estimate is conservative as far as uncertainties are concerned. These uncertainties cover the central values of the other two models,
 and they do not hinder the discrimination among different NP scenarios for $\sigma_0$, $\sigma_2$, $R_S$, $R_{T_t}$, $R_W$.

\section{$B_s$ decays}\label{app:f0etaetaprime}

We can consider $B_s\to f_0(980)\mu\mu$ and $B_s\to \eta^{(\prime)}\mu\mu$ decays following the same formalism as $B_d\to K_{(S,L)}\mu\mu$, up to a few changes:
\begin{itemize}
    \item the width difference $y\neq 0$ means that it becomes possible in principle to access $h_i$ coefficients,
    \item the $B_s$ mixing phase is much smaller than in the $B_d$ case and is thus competing with the decay phase from $V_{ts}$,
    \item the form factors describing these decays are not known as well as for $B\to K$ transitions, due to our limited knowledge of the structure of the $f_0$, $\eta$ and $\eta'$ mesons and their mixing with other states.
\end{itemize}
The expressions of the relevant form factors and matrix elements for $B_s\to\eta(')$ can be translated directly from the expressions in the $B_d\to K_S$ case described in this article.
The matrix elements relevant to $\bar B \to f_0$ transition can be defined as \cite{Colangelo:2010bg}
\begin{eqnarray}
\langle f_0(p) |\bar{s} \gamma_\mu \gamma_5 b|  \bar{B}_s(p_B)\rangle & =& -i \left[\left ( p_B + p \right )_\mu f_{+} (q^2) + \frac{m_{B_s}^2 - m_{f_0}^2}{q^2}q_\mu \left( f_0 (q^2) - f_{+} (q^2) \right) \right]\; , \nonumber \\
\langle f_0(p) |\bar{s}  \sigma_{\mu \nu} \gamma_5 b| \bar{B}_s(p_B)\rangle & =& - \left[ \left ( p_B + p \right )_\mu q_\nu -  \left ( p_B + p \right )_\nu q_\mu \right] \frac{f_T (q^2)}{m_{B_s} + m_{f_0}}  \; , \nonumber \\ %so multiplied by i
\langle f_0(p) |\bar{s}  \gamma_5  b| \bar{B}_s(p_B)\rangle & =& -i\frac{m_{B_s}^2 - m_{f_0}^2 }{m_b - m_s} f_0 (q^2) \; ,
\end{eqnarray}
leading to the amplitudes
\begin{eqnarray}
\HVK &=& i {\cal N}\frac{\sqrt{\KallenBs} }{2 \sqrt{q^2}} \left(\frac{2 m_b}{ m_{B_s} + m_{f_0} }  ( {\cal C}_{7} - {\cal C}_{7'}) f_T + ( {\cal C}_{9} - {\cal C}_{9'})   f_+  \right) \; ,  \nonumber \\
\HAK &=& i{\cal N}\frac{\sqrt{\KallenBs}  }{2 \sqrt{q^2}} ( {\cal C}_{10} - {\cal C}_{10'})  f_+ \; , \nonumber \\
\HSK &=& -i{\cal N} \frac{m_{B_s}^2-m_{f_0}^2}{2}\left(\frac{ ( {\cal C}_{S} - {\cal C}_{S'})}{m_b - m_s} \right)f_0 \; ,  \nonumber \\
\HPK &=&  -i{\cal N}\frac{m_{B_s}^2-m_{f_0}^2}{2}\left(\frac{ ( {\cal C}_{P} - {\cal C}_{P'})}{m_b - m_s} + \frac{ 2m_\ell}{ q^2  }   (- {\cal C}_{10} + {\cal C}_{10'}) \right)f_0 \; ,  \nonumber \\
\HTensK &=&   {\cal N}\frac{\sqrt{\KallenBs}  }{\sqrt{2} \left(  m_{B_s} + m_{f_0} \right)} \left({\cal C}_{T} + {\cal C}_{T'} \right) f_T \; ,  \nonumber \\
\HTenstK &=&  {\cal N}\frac{\sqrt{\KallenBs}  }{2 \left( m_{B_s} + m_{f_0} \right)} \left({\cal C}_{T} - {\cal C}_{T'} \right) f_T \; ,
\end{eqnarray}
where $\KallenBs\equiv \lambda(m_{B_s}^2, m_{f_0}^2, q^2)$. The amplitudes $h_X$ for $B_s\to f_0$ transitions can be obtained from $\bar{h}_X$ by taking the complex conjugate for all weak phases present in the amplitudes. 
We use the following form factors:
\begin{itemize}
    \item For $B_s\to \eta$ and $B_s\to\eta'$, we use the updated results of Refs.~\cite{Duplancic:2015zna}. 
    In particular, we use updated values~\cite{privatecommun}
   \begin{equation}
   \label{eq:updateval}\alpha_{B_s\eta}^+ =  0.5055  \pm 0.0195 \ , \quad\quad \alpha^+_{B_s \eta'} =  0.4928 \pm 0.0284 \ .
\end{equation}
We stress that the values for $\alpha^{0,T}$ in Ref.~\cite{Duplancic:2015zna,privatecommun} only include errors from varying $b_2$. However, currently full errors are not available. Note that also the central value for $\alpha^+_{B_s \eta'}$ in Eq.~\eqref{eq:updateval} from Ref.~\cite{privatecommun} differs from that quoted in Ref.~\cite{Duplancic:2015zna}. We stress that the $B_s \to \eta$ form factor at $q^2=0$, i.e. $f_i(0)$ is negative due to $\eta$-$\eta'$ mixing. 
    We are not aware of other determinations for these form factors.
    \item For $B_s\to f_0$, we use the Table I of Ref.~\cite{Colangelo:2010bg} as a reference.
    We take their results from Table I ($f_+= F_1$)
    with the parametrisation
\begin{equation}
    f_i(q^2) = \frac{f_i(0)}{1-a_i q^2/m_{B_s}^2 + b_i(q^2/m_{B_s}^2)^2} \ ,
\end{equation}
with $i=0,+, T$.
    We should however be careful that there is a large spread of the theoretical estimates of these form factors, as illustrated by Table 1 of Ref.~\cite{Cheng:2019tgh}. In case of asymmetric errors, we conservatively take the largest.
\end{itemize}

We can define the normalised angular observables:
\begin{equation}
    \sigma_i=\frac{s_i}{2\Gamma_\ell}\,,
    \qquad
    \theta_i=\frac{h_i}{2\Gamma_\ell}\,,
\end{equation}
considering also $h_i$ observables since $y\neq 0$. In principle, we could build equivalent quantities to $R_S$, $R_{T_t}$ and $R_W$ defined in \cref{eq:RS,eq:RTt,eq:RW} to isolate scalar and tensor amplitudes, but they would be affected by large uncertainties, as they would require dividing angular observables by $\sin \phi$ where $\phi$ is the small $B_s$ mixing angle. We will thus consider only $\sigma_i$ and $\theta_i$, which we provide in the various scenarios of interest in \cref{tab:3scen_eta,tab:3scen_eta_prime,tab:3scen_f_0} for the three $B_s$ decays, using our estimate (O.E.) for charm loops. 

Similarly to the $B_d\to K_S\ell\ell$ case, we can derive general relations in the case of real NP contributions to SM and chirally flipped Wilson coefficients. The amplitudes $h_X$ can be simplified exactly as in Sec.~\ref{sec:realNPSM}. If we take into account the different CKM coefficients (and different CP parities in some cases), we obtain for $B_s$ decays 
\begin{equation}
   \sigma_0=\sigma_2 = 0 \qquad 
   \theta_0 =-\theta_2 = -\frac{1}{2}\eta{(M)} \ ,
\end{equation} 
 up to $O(\lambda^2)$ corrections (were we neglect the lepton mass to obtain the expression of $\theta_2$).

The results illustrate the dependence of the angular observables on the NP scenario and the interest of measuring these quantities. Let us however that one should add to these results an additional theoretical systematic uncertainty reflecting our limited understanding of the form factors for these final states. Moreover, we neglected doubly Cabibbo-suppressed contributions with relative size of $O(\lambda^2)\simeq 4\%$, which are not included in our error estimates.

\begin{table}[t]
\begin{center}
\centering
  \begin{adjustbox}{max width=\textwidth}
\begin{tabular}{c|c|c|c|c|c|c}
  & SM & Scenario 1 & Scenario 2 & Scenario 3 & $C_S=0.2$ & $C_T=0.2$ \\ \hline
 $Br\times 10^8$ &$  4.8\pm 1.7 $&$  3.9\pm 1.4 $&$  3.4\pm 1.2 $&$  3.3\pm 1.2 $&$  4.8\pm 1.7 $&$  4.8\pm 1.7 $\\
 $F^{\ell }_H\times 10^2$ &$  2.47\pm 0.08 $&$  2.48\pm 0.09 $&$  2.49\pm 0.10 $&$  2.48\pm 0.09 $&$  2.51\pm 0.08 $&$  3.20\pm 0.26 $\\
 $A_0\times 10^2 $& 0 &$  0.0\pm 2.2 $&$  0.0\pm 0.4 $&$  0.\pm 4. $& 0 & 0 \\
 $\sigma _0\times 10^2$ &$  0.00\pm 0.05 $&$  -9.65\pm 0.70 $&$  -1.45\pm 0.16 $&$  -12.7\pm 1.9 $&$  0.00\pm 0.05 $&$  0.00\pm 0.05 $\\
 $\sigma _2\times 10^2$ &$  0.00\pm 0.06 $&$  9.46\pm 0.70 $&$  1.41\pm 0.15 $&$  12.4\pm 1.7 $&$  0.00\pm 0.06 $&$  0.00\pm 0.06 $\\
 $\theta _0\times 10$ &$  -5.00 \pm  0.00 $&$  -4.61\pm 0.02 $&$  -4.989\pm0.001 $&$  -3.70\pm 0.07 $&$  -5.00 \pm  0.00 $&$  -4.997\pm0.002 $\\
 $\theta _2\times 10$ &$  4.876\pm0.001 $&$  4.50\pm 0.02 $&$  4.864\pm0.005 $&$  3.61\pm 0.07 $&$  4.874\pm0.004 $&$  4.85\pm 0.01 $\\
\end{tabular}
\end{adjustbox}

\caption{Values of the observables in the SM, for the three different scenarios with new complex Wilson coefficients defined in~\eqref{eq:3scen} and for the scenarios with ${\cal C}_{S(T)}=0.2$. All quoted values are for $B_s\to f_0\mu\mu$ and are binned in $q^2$ over $[1,6]$ GeV$^2$. The inputs are taken from Table~\ref{tab:inputs}.
Neglected doubly Cabibbo-suppressed contributions are not included in our error estimates.}\label{tab:3scen_f_0}
\end{center}
\end{table}

\begin{table}[t]
\begin{center}
\centering
  \begin{adjustbox}{max width=\textwidth}
\begin{tabular}{c|c|c|c|c|c|c}
  & SM & Scenario 1 & Scenario 2 & Scenario 3 & $C_S=0.2$ & $C_T=0.2$ \\ \hline
 $Br\times 10^8$ &$  7.2\pm 1.2 $&$  5.8\pm 1.0 $&$  6.2\pm 1.0 $&$  6.3\pm 1.1 $&$  7.2\pm 1.2 $&$  7.2\pm 1.2 $\\
 $F^{\ell}_H\times 10^2$ &$  2.59\pm 0.02 $&$  2.63\pm 0.03 $&$  2.62\pm 0.03 $&$  2.57\pm 0.02 $&$  2.63\pm 0.03 $&$  3.22\pm 0.09 $\\
 $A_0\times 10^2$ & $0$ &$  0.0\pm 2.2 $&$  0.0\pm 1.2 $&$  0.\pm 4. $& $0$ & $0$ \\
 $\sigma _0 \times 10^2$ &$  0.00\pm 0.05 $&$  9.70\pm 0.60 $&$  -6.36\pm 0.26 $&$  -22.5\pm 0.6 $&$  0.00\pm 0.05 $&$  0.00\pm 0.05 $\\
 $\sigma _2 \times 10^2$ &$  0.00\pm 0.05 $&$  -9.47\pm 0.60 $&$  6.21\pm 0.27 $&$  22.0\pm 0.7 $&$  0.00\pm 0.05 $&$  0.00\pm 0.05 $\\
 $\theta _0 \times 10$ &$  5.00\pm  0.00 $&$  4.62\pm 0.02 $&$  4.87\pm 0.01 $&$  3.69\pm 0.08 $&$  5.00\pm 0.00 $&$  5.00\pm 0.00 $\\
 $\theta _2 \times 10$ &$  -4.871\pm0.001 $&$  -4.49\pm 0.02 $&$  -4.74\pm 0.01 $&$  -3.59\pm 0.08 $&$  -4.869\pm0.001 $&$  -4.844\pm0.003 $\\
\end{tabular}
\end{adjustbox}

% \begin{tabular}{c|c|c|c|c|c|c}
%   & SM & Scenario 1 & Scenario 2 & Scenario 3 & $C_S=0.2$ & $C_T=0.2$ \\ \hline
%  $Br\times 10^8$ &$  7.2\pm 1.2 $&$  5.8\pm 1.0 $&$  6.2\pm 1.0 $&$  6.3\pm 1.1 $&$  7.2\pm 1.2 $&$  7.2\pm 1.2 $\\
%  $F^{\ell}_H\times 10^2$ &$  2.585\pm 0.022 $&$  2.630\pm 0.025 $&$  2.617\pm 0.025 $&$  2.571\pm 0.020 $&$  2.630\pm 0.035 $&$  3.22\pm 0.09 $\\
%  $A_0\times 10^2$ & $0$ &$  0.0\pm 2.2 $&$  0.0\pm 1.2 $&$  0.\pm 4. $& $0$ & $0$ \\
%  $\sigma _0 \times 10^2$ &$  0.00\pm 0.05 $&$  9.70\pm 0.60 $&$  -0.636\pm 0.026 $&$  -2.25\pm 0.06 $&$  0.000\pm 0.005 $&$  0.000\pm 0.005 $\\
%  $\sigma _2 \times 10$ &$  0.000\pm 0.005 $&$  -0.95\pm 0.06 $&$  0.621\pm 0.027 $&$  2.20\pm 0.07 $&$  0.000\pm 0.005 $&$  0.000\pm 0.005 $\\
%  $\theta _0 \times 10$ &$  5.00000(0 \pm  4 ) $&$  4.616\pm 0.022 $&$  4.869\pm 0.008 $&$  3.69\pm 0.08 $&$  4.99534\pm 0.00032 $&$  4.9975\pm 0.0006 $\\
%  $\theta _2 \times 10$ &$  -4.8708\pm 0.0011 $&$  -4.494\pm 0.022 $&$  -4.741\pm 0.009 $&$  -3.59\pm 0.08 $&$  -4.8685\pm 0.0012 $&$  -4.8442\pm 0.0033 $\\
% \end{tabular}

\caption{Values of the observables in the SM, for the three different scenarios with new complex Wilson coefficients defined in~\eqref{eq:3scen} and for the scenarios with ${\cal C}_{S(T)}=0.2$. All quoted values are for $B_s\to \eta\mu\mu$ and are binned in $q^2$ over $[1,6]$ GeV$^2$. The inputs are taken from Table~\ref{tab:inputs}.
Neglected doubly Cabibbo-suppressed contributions are not included in our error estimates.}\label{tab:3scen_eta}
\end{center}
\end{table}

\begin{table}[t]
\begin{center}
\centering
  \begin{adjustbox}{max width=\textwidth}
\begin{tabular}{c|c|c|c|c|c|c}
  & SM & Scenario 1 & Scenario 2 & Scenario 3 & $C_S=0.2$ & $C_T=0.2$ \\ \hline
 $  Br\times10^8  $ &$  9.1\pm 1.9 $&$  7.4\pm 1.5 $&$  7.8\pm 1.6 $&$  8.1\pm 1.7 $&$  9.1\pm 1.9 $&$  9.2\pm 1.9 $\\
 $  F^{\ell }_H\times 10^2  $ &$  2.64\pm 0.04 $&$  2.70\pm 0.04 $&$  2.68\pm 0.04 $&$  2.63\pm 0.03 $&$  2.69\pm 0.04 $&$  3.29\pm 0.11 $\\
 $  A_0\times 10^2  $ & 0 &$  0.0\pm 2.3 $&$  0.0\pm 1.3 $&$  0.\pm 4. $& 0 & 0 \\
 $  \sigma_0\times10^2  $ &$  0.00\pm 0.05 $&$  9.7\pm 0.6 $&$  -6.3\pm 0.3 $&$  -22.5\pm 0.7 $&$  0.00\pm 0.05 $&$  0.00\pm 0.05 $\\
 $  \sigma_2\times10^2  $ &$  0.00\pm 0.05 $&$  -9.5\pm 0.6 $&$  6.2\pm 0.3 $&$  21.9\pm 0.6 $&$  0.00\pm 0.05 $&$  0.00\pm 0.05 $\\
 $  \theta_0\times10  $ &$  5.00 \pm 0.00 $&$  4.62\pm 0.02 $&$  4.87\pm 0.01 $&$  3.69\pm 0.08 $&$  5.00\pm 0.00 $&$  5.00\pm0.00 $\\
 $  \theta_2\times10  $ &$  -4.868\pm0.002 $&$  -4.49\pm 0.02 $&$  -4.74\pm 0.01 $&$  -3.59\pm 0.08 $&$  -4.867\pm0.002 $&$  -4.841\pm0.004 $\\
\end{tabular}
\end{adjustbox}
% \begin{tabular}{c|c|c|c|c|c|c}
%   & SM & Scenario 1 & Scenario 2 & Scenario 3 & $C_S=0.2$ & $C_T=0.2$ \\ \hline
%  $  Br\times10^8  $ &$  9.0\pm 1.9 $&$  7.3\pm 1.6 $&$  7.7\pm 1.7 $&$  7.9\pm 1.7 $&$  9.0\pm 1.9 $&$  9.0\pm 1.9 $\\
%  $  F^{\ell }_H\times 10^2  $ &$  2.671\pm 0.034 $&$  2.73\pm 0.04 $&$  2.71\pm 0.04 $&$  2.653\pm 0.031 $&$  2.72\pm 0.04 $&$  3.32\pm 0.11 $\\
%  $  A_0\times 10^2  $ & 0 &$  0.0\pm 2.1 $&$  0.0\pm 1.2 $&$  0.\pm 4. $& 0 & 0 \\
%  $  \sigma_0\times10  $ &$  0.000\pm 0.005 $&$  0.97\pm 0.06 $&$  -0.634\pm 0.027 $&$  -2.25\pm 0.06 $&$  0.000\pm 0.005 $&$  0.000\pm 0.005 $\\
%  $  \sigma_2\times10  $ &$  0.000\pm 0.005 $&$  -0.95\pm 0.06 $&$  0.619\pm 0.028 $&$  2.19\pm 0.07 $&$  0.000\pm 0.005 $&$  0.000\pm 0.005 $\\
%  $  \theta_0\times10  $ &$  5.00000(0 \pm 4) $&$  4.615\pm 0.022 $&$  4.869\pm 0.008 $&$  3.68\pm 0.08 $&$  4.9951\pm 0.0004 $&$  4.9973\pm 0.0007 $\\
%  $  \theta_2\times10  $ &$  -4.8665\pm 0.0018 $&$  -4.488\pm 0.022 $&$  -4.736\pm 0.009 $&$  -3.58\pm 0.08 $&$  -4.8641\pm 0.0019 $&$  -4.839\pm 0.004 $\\
% \end{tabular}

\caption{Values of the observables in the SM, for the three different scenarios with new complex Wilson coefficients defined in~\eqref{eq:3scen} and for the scenarios with ${\cal C}_{S(T)}=0.2$. All quoted values are for $B_s\to \eta^\prime\mu\mu$ and are binned in $q^2$ over $[1,6]$ GeV$^2$. The inputs are taken from Table~\ref{tab:inputs}. Neglected doubly Cabibbo-suppressed contributions are not included in our error estimates. }\label{tab:3scen_eta_prime}
\end{center}
\end{table}

\end{document}